\documentclass[showpacs,aps,preprint]{revtex4}
\usepackage{graphicx}
\usepackage{amsmath,amssymb}
\usepackage{accents}% necessary for widebar
\makeatletter
\def\widebar{\accentset{{\cc@style\underline{\mskip10mu}}}}
\makeatother

\def\gtsim{\mathrel{\hbox{\raise0.2ex
  \hbox{$>$}\kern-0.75em\raise-0.9ex\hbox{$\sim$}}}}
\def\ltsim{\mathrel{\hbox{\raise0.2ex
  \hbox{$<$}\kern-0.75em\raise-0.9ex\hbox{$\sim$}}}}

\def\mbold#1{\mbox{\boldmath $#1$}}

\def\mhalf{\mbox{$\frac{1}{2}$}}
\def\m3half{\mbox{$\frac{3}{2}$}}

\begin{document}

\title{
First order Coriolis-coupling for
rotational spectrum of a~tetrahedrally-deformed
core plus one-particle system
}

\author{Shingo Tagami and Yoshifumi R. Shimizu}
\affiliation{
 Department of Physics, Graduate School of Science,
 Kyushu University, Fukuoka 819-0395, Japan
}
\author{Jerzy Dudek}
\affiliation{
 \it Universit\'e de Strasbourg, CNRS, IPHC UMR 7178,
 F-67\,000 Strasbourg, France
}
\affiliation{
 \it Institute~of~Physics,~Marie Curie-Sk\l odowska University,
 PL-20\,031 Lublin, Poland
}

%\date{\today}

\begin{abstract}
The possible existence of shape-coexisting nuclear configurations with tetrahedral symmetry is receiving an increasing attention due to unprecedented nuclear structure properties, in particular in terms of the exotic 4-fold nucleonic level degeneracies and the expected long lifetimes which may become a new decisive argument in the exotic nuclei research programs. The present article addresses the rotational structure properties of the tetrahedrally-symmetric even-even core configurations coupled with a single valence nucleon. We focus on the properties of the associated Coriolis-coupling Hamiltonian proposing the solutions based on the explicit construction of the bases of the irreducible representations of the tetrahedral point-group on the one-hand side and the microscopic angular-momentum and parity projection nuclear mean-field approach on the other.
It is shown that for one-particle occupying an orbital belonging to
the $E_{1/2}$ or $E_{5/2}$ irreducible representation, the rotational
spectrum splits into two sequences, the structures analogous to those of the $K=1/2$ rotational bands in the axially symmetric nuclei.
Although the spectrum is generally more complicated
for one-particle occupying a 4-fold degenerate orbital belonging to the $G_{3/2}$ representation,
an appearance of the correlated double-sequence structures persists.
The spectra of the doubly-magic tetrahedral core plus one-particle systems can be well interpreted using the analytical solutions of the first order Coriolis-coupling Hamiltonian.
We introduce the notion of the generalized decoupling parameters, which determine the size of the energy-splitting between the double-sequence structures.
\end{abstract}

\pacs{21.10.Re, 21.60.Ev, 23.20.Lv}

\maketitle

%------------------------------------------------------------------------------
%------------------------------------------------------------------------------
\section{Introduction}
\label{sec:intro}

A great majority of atomic nuclei are non-spherical both in their ground-, and in the excited-states. This implies that their orientation in space can be defined and thus the corresponding systems may rotate collectively forming what is referred to as rotational bands. It turns out that the structure of the rotational bands and the related collective electromagnetic transitions depend on the geometrical symmetries of the nuclei in question and can be used for testing of the presence of certain point-group symmetries in nuclei.

The studies of the geometrical forms of nuclei found in the literature focus primarily on the quadrupole axial, in particular prolate and oblate shapes and their possible coexistence, and quadrupole triaxial ones; less frequently, on the octupole (pear-shape) deformations. The idea that the nuclear matter density in atomic nuclei may acquire more exotic symmetries resembling those of certain molecules was put forward  already in the 30's of the previous century in Ref.~\cite{JAW37}. It is natural to expect that nuclei in which tightly packed alpha-, and/or other light-clusters can coexist, may take more exotic symmetries and thus nearly at the same time, the alpha-cluster structures accompanied by a single-nucleon particle (hole) states have been discussed in Ref.~\cite{LRH38}. In particular, the structures composed of 4-, or 6 tightly-packed alphas become the prototypes of quantum systems, whose symmetry properties are governed by tetrahedral and octahedral point groups, and their associated the so-called double point-group realizations. At the same time the corresponding collective wave functions transform according to the irreducible representations of the point-groups in question. We return to the group-theory aspects in the more general context of non-alpha cluster nuclei in some detail later in this article.

Numerous studies of the nuclear alpha-cluster tetrahedral-symmetry prototype nucleus, $^{16}$O, have been undertaken in the past, cf.~early Refs.~\cite{DMD54,SLK56} -- and in many articles which followed. Specific efforts were undertaken later on to develop algebraic methods capable of describing the nuclear cluster structures, cf.~e.g.~Ref.~\cite{RBI00} focussing on the unitary groups and~Ref.~\cite{RBI02} discussing in particular the $D_{3h}$-symmetry. Interested reader may consult e.g.~Refs.~\cite{RBi15,RBi16,RBi17} and references therein, where the algebraic methods are applied in the context of various properties and observables in nuclei described within nuclear cluster structures. The most recent applications of these techniques in the context of the identification of the tetrahedral symmetry in $^{16}$O can be found in Refs.~\cite{RBi14,RBi17a}, cf.~also references therein.

Whereas proposing geometrical symmetries of the nuclear objects on the basis of the alpha (or for that matter any other light clusters) can be seen as a direct conceptual analogy with the molecular structures, finding such symmetries on the basis of the many-body (e.g.~mean-field) Hamiltonians is a totally different matter. Among early studies addressing the microscopic origin of the alpha-structures in nuclei by beginning the description with the one-particle (single-nucleon) wave functions such as the ones generated by a mean-field Hamiltonian while taking into account the model nucleon-nucleon interactions one finds Ref.~\cite{KWK58}. Our approach is relatively close to the nuclear mean-field description, which introduces explicitly the issue of the single-particle levels of the nucleonic spectra in the tetrahedral-symmetric mean-fields leading to a number of exotic nuclear structure properties. To give an example of such exotic properties let us recall that the tetrahedral-symmetry double point group, $T_d^D$, applicable to the mean-field Hamiltonians is characterized by two two-dimensional irreducible representations and one 4-dimensional one. This implies that certain nucleonic levels in the tetrahedral symmetry nuclei should produce a very exotic, so far unprecedented feature: some of the levels may be occupied by up to 4 nucleons of the same isospin. Thus nuclei obeying tetrahedral symmetry exactly may, among other exotic features, manifest the presence of the 16-fold degenerate particle hole excitations. One of the early predictions focusing on the 4-fold degeneracies in realistic mean-field calculations for heavy nuclei can be found in Ref.~\cite{XLD94}. The mechanisms involving the presence of highly degenerate excited states propagate in an interesting manner to the rotational properties of the systems composed of particles coupled to the collective rotors and the underlying so-called Coriolis-coupling mechanism. This mechanism will be addressed explicitly in the present article in the case of the tetrahedral-symmetry quantum rotors.

Let us mention in passing another of those exotic symmetry properties which makes the whole matter particularly interesting for the international programs of the exotic nuclei research. Indeed it can be shown that nuclei with the exact tetrahedral symmetry produce neither collective $E1$-, nor $E2$-transitions, the corresponding multipole dipole and quadrupole moments vanishing due to symmetry hindrance. Such a hindrance is expected to lead to an increase in the lifetimes of such exotic states by several orders of magnitude making them particularly attractive in the research of the exotic nuclei in which tetrahedral symmetry isomers may live significantly longer than e.g. the nuclear ground states. All these features attracted particular attention within the nuclear mean-field community. In particular, one of the moderately heavy (non-alpha-cluster) nuclei in which the presence of tetrahedral symmetry has been predicted by independent teams of researchers working with the self-consistent Hartree-Fock-Bogolyubov method is $^{80}_{40}$Zr$^{}_{40}$ nucleus as early as towards the end of the previous century, Refs.~\cite{STK98,KMY01} and later on, Refs.~\cite{KZb06,TSD15}. Later on several quantum mechanisms and their description pertinent to studying the point-group symmetries in nuclei have been developed. This concerns in particular: constructing the nuclear mean-field Hamiltonians with a predefined point group symmetry, relating systematically the Hamiltonian-symmetry groups and nuclear stability, constructing quantum rotor collective model-Hamiltonians of predefined point-group symmetry, multi-dimensional deformation spaces involving in particular the so-called isotropy groups and orbits, detailed analysis of the transformations between the laboratory and rotating frames and the associated symmetrization group, and several others. The reader interested in these issues can consult an overview article Ref.~\cite{JDu13}, cf.~also Ref.~\cite{DGM10}.

The most recent discussion of a new approach to examining the experimental evidence for the presence of the tetrahedral and octahedral symmetries in nuclei focussed on the realistic example of $^{152}$Sm can be found in Ref.~\cite{DCD18}.

The nuclear tetrahedral symmetry invokes an extra stability leading to
the so-called tetrahedral magic numbers.
We have performed the angular-momentum and parity projection calculation
from the tetrahedrally deformed mean-field states~\cite{TSD13,TSD15},
and found that the characteristic spectra suggested by the group theory
naturally come out for even-even closed tetrahedral-shell nuclei
by such a microscopic approach.
In the present work, we extend this type of research for nuclei
with a valence nucleon on top of a doubly closed tetrahedral-shell configuration at the asymptotic limit of very large tetrahedral deformations.

For the axially-symmetric quadrupole-deformed nuclei,
the effect of an odd nucleon on the collective rotation
is well-known and described in terms the quantum analogue of the Coriolis interaction,
see e.g.~Ref.~\cite{BM75,RS80}.
In the present article we choose to follow, in analogy,
the first-order Coriolis-coupling description
for the strongly-deformed systems
with the tetrahedral point-group symmetry by employing the techniques of the group representation theory, see below.
For this purpose, the wave function of
the so-called strong-coupling type~\cite{BM75},
which is suitable for large deformation, is introduced.
It is found that the matrix elements of the first-order Coriolis-coupling
can be diagonalized analytically
and formula for the rotational excitation-energy spectrum can be derived.
We present the results of the microscopic projection calculations and show that they
can be interpreted in terms of the generalized {\em decoupling parameter(s)}
in analogy to the axially-symmetric quadrupole deformation.

The paper is organized as follows.
We present how the Coriolis coupling
can be calculated in Sec.~\ref{sec:formulation},
where the necessary mathematical ingredients are included
with the help of group theory.
In Sec.~\ref{sec:results} we present the results of energy spectra
for the typical core plus one-particle system in $^{81}$Zr nucleus,
where the microscopic angular-momentum projection method is employed
with the Hamiltonian composed of
the Woods-Saxon mean-field and the schematic interactions~\cite{TS12}.
The results are investigated in relation to the energy expression
obtained by the calculation of the Coriolis coupling.
Sec.~\ref{sec:concl} is devoted to the summary and conclusions.
Some mathematical details are discussed in Appendices.
Preliminary results were already published in Ref.~\cite{TSF14}.

%------------------------------------------------------------------------------
%------------------------------------------------------------------------------

\section{First order Coriolis-coupling for tetrahedrally-deformed systems}
\label{sec:formulation}

In the present work, we formulate the generalized decoupling parameter technique known from the traditional literature describing the coupling of an odd particle  with a quadrupole-deformed second-order quantum-rotor. A discussion of the structure of the Hamiltonian of such systems, in the form of the so-called particle-rotor model, can be found for instance in Sec.~4-2 of Ref.~\cite{BM75} or in Sec.~3.3 of Ref.~\cite{RS80}. The Hamiltonian in question has the general form
\begin{align}
       \hat H= \hat H_{\rm mf}+\hat H_{\rm rot},
                                                      \label{eq:Htot}
\end{align}
where the first term is a deformed nuclear mean-field Hamiltonian and the second one describes the collective rotation of the system.

The generalization considered in this article consists in obtaining a mathematically similar decoupled-band picture for the systems with tetrahedral symmetry rather than triaxial or axial ellipsoids. To introduce the framework of the presentation we first discuss in some detail the structure of the Hamiltonian applied here.

The first term in Eq.~(\ref{eq:Htot}) represents the mean-field Hamiltonian, which is assumed to be invariant under the symmetry elements of the tetrahedral point-group, thus in general breaking the symmetry under inversion. In the following we will work under the approximation of no residual interaction included in the Hamiltonian, thus in particular ignoring the nuclear pairing. Such an approximation is partially justified by the fact that tetrahedral-symmetry nuclear-configurations are due to relatively large tetrahedral shell-gaps, the mechanism known to weaken the pairing interactions represented  by the so called BCS-$\Delta$ gap-parameter. Moreover, the presence of an odd nucleon weakens the pairing interactions even more due to the well known blocking mechanism.

The second term represents the quantum-rotor Hamiltonian. In the present work we choose a quadratic form involving the three components of the collective angular momentum operator $\hat{R}\equiv\{\hat{R}_1,\hat{R}_2,\hat{R}_3\}$,
\begin{align}
  \hat H_{\rm rot}=\sum_{i=1}^3\frac{{\hat R_i}^2}{2 {\cal J}_i},
                                                      \label{eq:Hrot}
\end{align}
but interested reader may consult alternative formulations which can be found in the literature, cf.~Ref.~\cite{JDG01}.
Here and in what follows we use the body-fixed coordinate frame,
and the quantities, ${\cal J}_1$, ${\cal J}_2$, ${\cal J}_3$,
are the moment of inertia around the three principal axes.
The total angular-momentum operator $\mbox{\boldmath $\hat I$}$ is composed
of the rotor collective angular momentum $\mbox{\boldmath $\hat R$}$ and
of the valence-particle angular momentum~$\mbox{\boldmath $\hat \jmath$}$ contributions:
\begin{align}
       \mbold{\hat I}=\mbold{\hat R}+\mbold{\hat \jmath},
                                                      \label{eq:IRj}
\end{align}
and it follows that the rotor Hamiltonian can be written down as
\begin{align}
 \hat H_{\rm rot}=\hat H_{\rm coll}+\hat H_{\rm rec}+\hat H_{\rm cor},
                                                      \label{eq:Hcomp}
\end{align}
with
\begin{align}
  \hat H_{\rm coll}=\sum_{i=1}^3\frac{{\hat I_i}^2}{2{\cal J}_i},\quad
  \hat H_{\rm rec}=\sum_{i=1}^3\frac{{\hat \jmath_i}^2}{2{\cal J}_i},\quad
  \hat H_{\rm cor}=-\sum_{i=1}^3 \frac{{\hat I_i}{\hat \jmath_i}}{{\cal J}_i}.
                                                      \label{eq:Heach}
\end{align}
Above, $\hat H_{\rm coll}$ describes the collective rotational energy,
and the second term, $\hat H_{\rm rec}$, represents the so-called recoil energy of the valence particle. Some authors use the argument that this latter term, which depends only on the intrinsic degrees of freedom, can be absorbed in the mean field part of the Hamiltonian and, assuming that the corresponding modifications of the mean field are small, its presence is neglected. Other authors, arguing that the most often used mean fields do not contain the necessary framework allowing to include the recoil-term, and calculate the corresponding impact explicitly using alternative approaches, cf.~e.g.~Ref.~\cite{ORG75}, or deepen the detailed description involving the two-body mechanisms of the corresponding over-all effect as e.g.~Ref.~\cite{REO79} and/or employ the links with other excitation modes as e.g.~scissor-mode, cf.~Ref.~\cite{SI92}. These early studies were followed by more recent ones but since in the present article we neglect this term as an approximation, we do not address these issues anymore. The last, so-called Coriolis-coupling term
between the total system and the valence particle,
$\hat H_{\rm cor}$, will be explicitly treated in the present work.

In the following we use the $\hbar=1$ unit if not stated otherwise.

%------------------------------------------------------------------------------
\subsection{The case of axial symmetry: splitting of $\mbold{K=1/2}$ rotational bands}
\label{sec:axsym}

Let us begin by recalling the axially symmetric case
with the $z$ (or 3-rd) axis chosen as the symmetry axis in the body-fixed frame.
The eigenvalue $K$ of angular momentum $\hat I_3$, which coincides with
the eigenvalue of $\hat \jmath_3$, is a good quantum number;
see, e.g., Secs.~4-2 and~4-3 of Ref.~\cite{BM75}
or Sec.~3.3.1 of Ref.~\cite{RS80}.
With the requirement of the ${\cal R}$-invariance
(here the signature ${\cal R}$ is the operation of rotation through $\pi$ about the $y$-axis
in the body-fixed frame, see Sec.~\ref{sec:doublex} for details),
the Coriolis-coupling effect for such a system
can be easily calculated;
the leading-order expression for the rotational excitation-energy spectrum
is given by
Eq.~(4-61) in Sec.~4-3a of Ref.~\cite{BM75}, i.e.,
\begin{align}
    E_{K}(I)=\frac{1}{2{\cal J}}\big[ I(I+1) -K^2
           + a\, (-1)^{I+1/2}(I+1/2)\delta_{K,1/2} \big],
                                                   \label{eq:Edecax}
\end{align}
where ${\cal J}\equiv {\cal J}_1={\cal J}_2$.
Thus, for $K=1/2$ rotational band,
the spectrum splits into two sequences because of the oscillations of the second term
\begin{align}
      (-1)^{I+1/2}(I+1/2)
      =
      \biggl\{
      \begin{matrix}
      -(I+\mhalf),
      & I=\mbox{even integer}+\mhalf, \cr
     \ \ (I+\mhalf), & I=\mbox{even integer}-\mhalf.
     \end{matrix}
     \biggr.
                                                    \label{eqn.07}
\end{align}
The size of splitting is determined by the
so-called decoupling parameter
\begin{align}
 a\equiv-\langle \phi_{K=1/2} | \hat \jmath_+  e^{i\pi \hat \jmath_2}|
  \phi_{K=1/2} \rangle,
                                                   \label{eq:aaxK12}
\end{align}
where $| \phi_{K=1/2} \rangle$ is the axially deformed single-nucleon wave function of a valence particle.
Note that the splitting of rotational energy spectrum appears only
for the $K=1/2$ band in the axially-symmetric rotor
(for the band with $K>1/2$ the Coriolis-coupling effect is of higher order).

In the following, we will see that
in the case of tetrahedral deformation there is always $K$-mixing
and the Coriolis coupling is effective for
all the rotational bands in the core plus one-particle systems.
It will be further shown that the similar energy expression and
the splitting to two rotational sequences are obtained
by the Coriolis coupling with slightly different definition
of the ``decoupling parameter(s)''.

%------------------------------------------------------------------------------
\subsection{Strong-coupling limit for the wave functions in the presence of a point-group symmetry}
\label{sec:scwf}

The eigenstates $|I M K\rangle$ of the axially-symmetric collective-rotor Hamiltonian involve $(I,M)$, the eigenvalues of angular-momentum and its third projection in the laboratory frame,
and $K$, the eigenvalue of $\hat I_3$ in the body-fixed reference frame.
These eigenstates can be taken as Wigner $D$-functions,
$D^{I*}_{MK}(\omega)$,
depending on the Euler angles $\omega=(\alpha,\beta,\gamma)$.
We follow the convention of Ref.~\cite{Ed57}
for the angular-momentum algebra in the present work.
When analyzing systems with point-group symmetries, however,
a complication arises since the constructed wave functions
should transform as irreducible representations of the considered point group $G$ -- in our case tetrahedral. We say that each wave-function belongs to an irreducible representation of $G$.

Irreducible representations of the tetrahedral group will be labelled
with symbol $\lambda$; each irreducible representation
is characterized by its dimension, $f_\lambda$.
We introduce an extra quantum number $\mu$ to distinguish between
various basis states belonging to the same representation~$\lambda$
(anticipating the results of the discussion below a convenient choice of the quantum number~$\mu$ in the tetrahedral symmetry case
will the so-called $z$-doublex quantum number defined in Sec.~\ref{sec:doublex}).
Collective wave-functions respecting the discussed
point-group symmetry can be written down as
\begin{align}
      |I^\pi M\lambda\mu\beta\rangle
      =
      \sum_K|I^\pi MK\rangle \,
      C^\pi_{IK,\lambda\mu\beta} , \quad
      (\mu=1,\cdots,f_\lambda;\,\beta=1,\cdots,n^{\lambda}_{I\pi}),
                                                   \label{eq:lambdawf}
\end{align}
where we also introduced the parity quantum number $\pi=\pm 1$,
and $\beta$ is an additional quantum number
necessary to specify the point-group symmetric state
with angular-momentum and parity $I^\pi$,
whose occurrence numbers, $n^{\lambda}_{I\pi}$
can be found in the literature, cf.~e.g., Table VI and VIII in Appendix of Ref.~\cite{TSD13}. The expansion coefficients $C^\pi_{IK,\lambda\mu\beta}$ are for the moment unknown and will be specified later.

For the core plus one-particle systems,
the intrinsic single-nucleon states are described by the eigenstates of the deformed mean-field Hamiltonian.
These eigenstates will be denoted as $|\phi_{\lambda\mu}\rangle$ since they should transform according to the irreducible representations ($\lambda\mu$) of the same point-group.
For the sake of the following discussion it will be possible to omit other quantum numbers characterizing the single-nucleon properties.
For sufficiently large deformations,
the following ``strong-coupling'' wave function structure
\begin{align}
 |\Psi^{\lambda}_{I^\pi M\beta} \rangle=
 \frac{1}{\sqrt{f_\lambda}}
 \sum_{\mu=1}^{f_\lambda} |I^\pi M \lambda \mu \beta \rangle
  | \phi_{\lambda \mu} \rangle
                                                   \label{eq:strcwf}
\end{align}
is expected to be a good approximation~\cite{BM75}.

The collective and the intrinsic wave functions,
$|I^\pi M \lambda \mu \beta \rangle$ and $| \phi_{\lambda \mu} \rangle$,
should have consistent transformation properties in the sense that whereas the collective part transforms according to the representation here denoted as $\hat D_{\rm e}(g)$
\begin{align}
       \hat D_{\rm e}(g)|I^\pi M \lambda \mu \beta \rangle
       =
       \sum_{\mu'} |I^\pi M \lambda \mu' \beta \rangle
                    D^{[\lambda]}_{\mu\mu'}(g),
                                                   \label{eq:lrepe}
\end{align}
the intrinsic (single-nucleon) wave functions transform according to representation $\hat D_{\rm i}(g)$,
\begin{align}
       \hat D_{\rm i}(g)|\phi_{\lambda\mu} \rangle
       =
       \sum_{\mu'}|\phi_{\lambda\mu'} \rangle D^{[\lambda]}_{\mu'\mu}(g),
                                                   \label{eq:lrepi}
\end{align}
for an arbitrary symmetry-group element $g \in G$.
Operators $\hat D_{\rm i}(g)$ and $\hat D_{\rm e}(g)$ are
the group-representation operators acting in the spaces of intrinsic and collective
wave-functions, respectively, and $D^{[\lambda]}_{\mu\mu'}(g)$ is
the common unitary matrix for each group element $g$ in the irreducible representation $\lambda$ (observe different orders of the indices $\mu$ and $\mu'$ in Eqs.~(\ref{eq:lrepe}) and (\ref{eq:lrepi})).
In other words, $\hat D_{\rm e}(g)$ transforms the collective states
in the same way
as $\hat D_{\rm i}^{(c)}(g^{-1})$ transforms the intrinsic single particle states, where $\hat D_{\rm i}^{(c)}$ is
the complex conjugate representation of $\hat D_{\rm i}$, cf.~Ref.~\cite{Ham62}.

The transformation operators for $g \in G$ are given explicitly by
\begin{align}
      \hat D_{\rm i}(g)
      =
      \hat \Pi_{\rm i}(g)\
      e^{i\gamma(g) \hat \jmath_3}
      e^{i\beta(g) \hat \jmath_2}
      e^{i\alpha(g) \hat \jmath_3}
                                                   \label{eq:topi}
\end{align}
and
\begin{align}
       \hat D_{\rm e}(g)
       =
       \hat \Pi_{\rm e}(g)\
       e^{i\alpha(g) \hat I_3}
       e^{i\beta(g) \hat I_2}
       e^{i\gamma(g) \hat I_3}.
                                                   \label{eq:tope}
\end{align}
Above, $\alpha(g)$, $\beta(g)$ and $\gamma(g)$ are Euler angles
corresponding to the discrete rotations represented
by $g\in G$, and $\hat \Pi_{\rm i}(g)=\hat \pi$,
the operator of inversion in the intrinsic reference frame
if $g$ contains inversion, alternatively $\hat \Pi_{\rm i}(g)=1$. Operator
$\hat \Pi_{\rm e}(g)$ is defined in full analogy but for the
collective degrees of freedom.
Note that the rotation operators for the collective and intrinsic
degrees of freedom are formally different
since the angles $\alpha$ and $\gamma$ are interchanged.
This is a consequence of the fact that the components of
$\mbold{\hat \jmath} \equiv \{\hat \jmath_1, \hat \jmath_2, \hat \jmath_3\}$
obey the usual
commutation relations of the form
$[\hat{\jmath}_1,\hat{\jmath}_2]=i\,\hat{\jmath}_3$ etc,
whereas the components of $\mbold{\hat I} = \{\hat I_1, \hat I_2, \hat I_3\}$ satisfy the analogous commutation relation but with opposite signs on the right-hand sides.
It follows that
$_{\rm i}\langle IK'| e^{i\gamma \hat \jmath_3} e^{i\beta \hat \jmath_2}
e^{i\alpha \hat \jmath_3}|IK\rangle_{\rm i}=
_{\rm e\!\!}\langle IK| e^{i\alpha \hat I_3} e^{i\beta \hat I_2}
e^{i\gamma \hat I_3}|IK'\rangle_{\rm e}$ and
in the following we omit the subscript ``${\rm i}$'' or ``${\rm e}$''
as long as there is no risk of confusion.

For the transformations of the rotor-associated functions
we introduce operators $\hat D_{\rm r}(g)$ identical to
$\hat D_{\rm e}(g)$ since the components of
$\mbold{\hat R} = \{\hat R_1, \hat R_2, \hat R_3\}$ satisfy the same commutation relations as those of $\mbold{\hat I}$.
We may straightforwardly verify that using
$\hat \Pi_{\rm r}(g)
 =
 \hat \Pi_{\rm e}(g)\hat \Pi_{\rm i}(g)
$
and
$
 \mbold{\hat R}
 =
 \mbold{\hat I}-\mbold{\hat \jmath},
$
one obtains
\begin{align}
       \hat D_{\rm r}(g)
       =
       \hat D_{\rm e}(g) \,
       \hat D_{\rm i}(g^{-1}),
                                                  \label{eq:Drei}
\end{align}
and it follows that the wave function in Eq.~(\ref{eq:strcwf}) is invariant
under $\hat D_{\rm r}(g)$,
\begin{align}
       \hat D_{\rm r}(g)|\Psi_{I^\pi M \beta} \rangle
       =
       |\Psi_{I^\pi M \beta} \rangle,\quad g \in G .
                                                  \label{eq:Symwf0}
\end{align}
Alternatively,
\begin{align}
       \hat D_{\rm e}(g)|\Psi_{I^\pi M \beta} \rangle
       &=
       \sum_{\mu\mu'}|I^\pi M \lambda \mu \beta \rangle
       |\phi_{\lambda\mu'} \rangle D^{[\lambda]}_{\mu'\mu}(g)
                                                  \nonumber \\
       &=
       \hat D_{\rm i}(g)|\Psi_{I^\pi M \beta} \rangle,\quad g \in G.
                                                  \label{eq:Symwf}
\end{align}
We say that the results of transformations of the collective wave functions and
those the intrinsic variables are conjugated,
which is indeed the required symmetry property
with respect to the point-group $G$
(see e.g.~Sec.~4-2c of Ref.~\cite{BM75}).

%------------------------------------------------------------------------------
\subsection{Coriolis coupling for tetrahedrally-deformed
core plus one-particle system}
\label{sec:coritetra}

To discuss the spectra for the even-even core plus one-particle systems
generated by the tetrahedral-symmetric Hamiltonian,
we will introduce three irreducible representations
of the $T_d^D$ group known in literature, cf.~Secs.~9-6 and~9-7
of Hamermesh, Ref.~\cite{Ham62}.
We use here the notation as in Table VIII, Appendix of Ref.~\cite{TSD13}
according to which we set
$\lambda=E_{1/2}$, $E_{5/2}$ and $G_{3/2}$ for the representations
denoted as $E'_1$, $E'_2$ and $G'$ in the above textbook.
The $E_{1/2}$ and $E_{5/2}$ orbitals are 2-fold degenerate,
while the $G_{3/2}$ orbital is 4-fold degenerate.
The irreducible representations appropriate for the boson-like tetrahedral $T_d$-symmetric even-even systems are denoted according to the same references as $A_1$, $A_2$, $E$, $F_1$ and $F_2$. In the ground-state of an even-even core nucleus, all the 2- and 4-fold degenerate single-particle orbitals are fully occupied forming an $I^\pi=0^+$ configuration. Such a state may belong exclusively to the $A_1$ irreducible representation. It then follows that the single-particle state of the odd, valence nucleon coupled to the ground-state, determines uniquely
the representation of the total odd-$A$ system.

Since the classical tetrahedral symmetric bodies have all the three principal-axis
moments of inertia equal, ${\cal J}_1={\cal J}_2={\cal J}_3\equiv {\cal J}$,
we impose this result in the rotor Hamiltonian in Eq.~(\ref{eq:Hrot}).
Then, the total rotational energy described by
$H_{\rm coll}$ in Eq.~(\ref{eq:Heach}) is given by the usual quadratic
spin dependence, $E_{\rm coll}=I(I+1)/2{\cal J}$.
In order to obtain the spectra for the core plus one-particle systems,
one has to diagonalize the first-order Coriolis-coupling Hamiltonian,
\begin{align}
       \langle \Psi^{\lambda}_{I^\pi M \beta'} |
       \hat H_{\rm cor} |\Psi^{\lambda}_{I^\pi M \beta} \rangle
       =
       -\frac{1}{{\cal J} }\,
       \frac{1}{f_\lambda}\sum_{\mu \mu'}
       \langle I^\pi M \lambda \mu' \beta'|
       \mbold{\hat I} | I^\pi M \lambda \mu \beta \rangle
       \cdot
       \langle \phi_{\lambda \mu'} |
       \mbold{\hat \jmath} | \phi_{\lambda \mu}\rangle .
                                                    \label{eq:Hcor}
\end{align}
In the following we discuss diagonalization of this coupling matrix analytically
by suitably constructed basis states,
which can be performed exactly for $\lambda=E_{1/2}$ and $E_{5/2}$
and approximately for $\lambda=G_{3/2}$.

Because the quantum number $M$ does not play any dynamical role
for the energy spectra,
we omit it
to simplify the notation.

%------------------------------------------------------------------------------
\subsection{Doublex symmetry and the corresponding good quantum number}
\label{sec:doublex}

In the following, we consider the tetrahedral group, $G=T_d$,
and the tetrahedral double group, $G^D=T_d^D$.
We will begin by specifying the body-fixed coordinate frame.
For this purpose we will introduce the nuclear surface parameterization in terms of spherical harmonics
\begin{align}
       R(\theta,\varphi)
       \propto
       \bigl[1+\sum_{lm}\alpha^*_{lm}Y_{lm}^{}(\theta,\varphi)\bigr],
                                                    \label{eq:radR}
\end{align}
and use coordinate system for which the lowest order tetrahedral-deformed shapes are described by $\alpha_{32}=\alpha_{3-2}$, see e.g.~Ref.~\cite{DDD07}.

In analyzing rotational properties of nuclei whose shapes are described
in terms of the spherical harmonics the discrete symmetries referred to
as $y$-signature and $y$-simplex turned out to be very practical.
They are defined in a {\em body-fixed reference frame} as the operations of rotation through the angle of $\pi$ about $y$-axis,
$
 \hat{\mathcal{R}}_y
 \equiv
 \exp(i \pi \hat{J}_y) \,(\equiv \hat R_y\mbox{ below})
$
and a combination of the latter with the operation of inversion,
$
 \hat{\mathcal{S}}_y
 \equiv
 \hat{\mathcal{R}}_y\,\hat{\pi} \,(\equiv \hat S_y\mbox{ below}),
$
respectively.  In analogy one may introduce another useful discrete operation
referred to as doublex, cf.~e.g.~Refs.~\cite{SDF05,JDu13} by
$
 \hat{\mathcal{D}}_y
 \equiv
 \exp[i(\pi/2) \hat{J}_y]\,\hat{\pi}.
$
In what follows it will be more practical to work with the $z$-doublex,
$
 \hat{\mathcal{D}}_z
 \equiv
 \exp[i (\pi/2) \hat{J}_z]\,\hat{\pi} \,(\equiv \hat{S}^z_4\mbox{ below}).
$
Here and below we use $(\hat J_x,\hat J_y,\hat J_z)$ as generic symbols
representing angular-momentum operators with the following correspondence
\begin{align}
       (\hat J_x,\hat J_y,\hat J_z)
       \leftrightarrow
       (\hat \jmath_1,\hat \jmath_2,\hat \jmath_3),
       \quad
       \mbox{or}
       \quad
       (\hat J_x,\hat J_y,\hat J_z)
       \leftrightarrow
       (\hat I_1,-\hat I_2,\hat I_3), ~ {\rm etc.}
                                                   \label{eq:JcorjI}
\end{align}
For even systems of fermions we have
\begin{align}
      \hat{\mathcal{D}}^{\,4}_z=1 \;\;\to\;\; d_z^{\,4}=1,
                                                        \label{eq:Dbleig}
\end{align}
and following Ref.~\cite{SDF05} the eigenvalues $d_z$ of $\hat{\mathcal{D}}_z$
can be written down as
$d_z=e^{i\,\pi\delta}$ where the fourth-order roots can be parametrized with the help of $\delta =0, \frac{1}{2}, 1, \frac{3}{2}$. Any value of $\delta$ differing from the above values by an integer multiple of $2$ will be equivalent to one of the above. In what follows we will be using the doublex exponent (an analogue of the signature exponent) denoted as $\mu$; we have the correspondence $\mu \leftrightarrow 2 \delta$ and because of the presence of the factor of $\pi/2$ in the exponential in the definition of doublex operation, the physically significant values of $\mu$ can be determined modulo 4.

Irreducible representations $\lambda$ will be used for
examining the properties of either the collective
or the intrinsic wave functions. It will be instructive
to introduce certain formal
properties of the basis states $|\lambda\mu \rangle$
of the representation $\lambda$. Since for
$G=T_d$ and $T_d^D$ groups, doublex operation
associated with the $z$-axis
is among the symmetry elements,
it will be possible to choose
the quantum number $\mu$ in Eq.~(\ref{eq:lambdawf})
for parametrizing its eigenvalues as follows
\begin{align}
       \hat S^z_4 |\lambda\mu \rangle
       =
       e^{i \frac{\pi}{2} \mu} |\lambda\mu \rangle,
      \quad
      \hat S^z_4
      \equiv
      \hat \Pi\, e^{i \frac{\pi}{2} \hat J_z}.
                                                   \label{eq:zdblx}
\end{align}
Thus, for the general angular-momentum eigenstate $|I^\pi K\rangle$
in a body-fixed reference frame, where $K$ represents the 3rd (or $z$)-component of the angular-momentum and
the parity $\pi=\pm 1$ (distinction should be made between two different roles of the symbol $\pi$ in the following expression),
the $z$-doublex(-exponent) $\mu$ is given by
\begin{align}
       \hat S^z_4 |I^\pi K \rangle
       =
       \pi e^{i \frac{\pi}{2} K} |I^\pi K \rangle
       \quad
       \Rightarrow \quad
       \mu=K+1-\pi \  \ (\mbox{mod}\ 4),
                                                   \label{eq:jmdblx}
\end{align}
where $d_z=\pi e^{i \frac{\pi}{2} K}$ represents the doublex eigenvalue.

It follows that $\mu=0, \pm 1, 2$ (mod~4), for boson systems ($T_d$),
and $\mu=\pm \frac{1}{2}, \pm \frac{3}{2}$ (mod~4),
for fermion systems ($T_d^D$),
and it is easy to find the appropriate values of $z$-doublex exponent in each representation.
They are collected in Table~\ref{tab:dblx}
(see Appendix~\ref{sec:DblxSy} for details).

\begin{table*}[hbtp]
\begin{tabular}{c|ccccc|ccc}
 & & & $T_d$ & & & & $T_d^D$ & \\
 \hline
$\lambda\ $ & $\ \ A_1\ $ & $\ A_2\ $ & $\ E\ $ & $F_1(T_1)$ & $F_2(T_2)\ $ &
  $\ E_{1/2}(E'_1)$ & $\ E_{5/2}(E'_2)$ & $\ G_{3/2}(G')$ \\
 \hline
$f_\lambda\ $ & $\ 1$ & $1$ & $2$ & $3$ & $3$ &
  $2$ & $2$ & $4$  \\
$\mu\ $ & $\ 0$ & $2$ & $\ 0,2\ $ & $\ 0,\pm 1\ $ & $\ \pm 1,2\ $ &
 $\pm \frac{1}{2}$ & $\pm \frac{3}{2}$ & $\ \pm \frac{1}{2},\pm \frac{3}{2}$  \\
\end{tabular}
\caption{The dimension $f_{\lambda}$ of
 the irreducible representations in the $T_d$ and $T_d^D$ groups
 and the values of $z$-doublex exponent $\mu$ associated with them.
}
                                                    \label{tab:dblx}
\end{table*}

Let us notice that the $y$-simplex operation $\hat S_y$ introduced above,
which is a group element of both $T_d$ and $T_d^D$,
satisfies
\begin{align}
      \hat S_y^{\dagger} \hat S_4^z \hat S_y
      =
      {\hat S_4^z}^{\dagger},
      \quad
      \hat S_y
      \equiv
      \hat \Pi \hat R_y,
      \quad
      \hat R_y\equiv e^{i \pi \hat J_y},
                                                    \label{eq:zdblsig}
\end{align}
and it follows that the operation of $\hat S_y$ changes
the $z$-doublex from $\mu$ to $-\mu$ (mod~4). Therefore,
\begin{align}
      \hat S_y | \lambda\mu \rangle
      \propto
      | \lambda-\mu \rangle \,
      \quad\mbox{ for } \mu \ne 0,2,
                                                    \label{eq:Sybasis}
\end{align}
and for the states with $\mu=0$ and $2$ we arrive at an extra symmetry,
for which $(\hat S_y)^2|\lambda\mu\rangle=|\lambda\mu\rangle$. Consequently
\begin{align}
      \hat S_y | \lambda\mu \rangle
      =
      \pm | \lambda\mu \rangle
      \quad\mbox{ for } \mu=0,2 \,.
                                                    \label{eq:Sysym}
\end{align}
The signs of simplex,
i.e., symmetry or anti-symmetry with respect to $\hat S_y$,
for all possible representations having $\mu=0$ and $2$
are summarized in Table~\ref{tab:Sysym}
(see Appendix~\ref{sec:DblxSy} for details).

\begin{table*}[hbtp]
\begin{tabular}{c|cccccccc}
$\lambda\mu\ $ &
$\ A_10\ $ & $\ A_22\ $ & $\ E0\ $ & $\ E2\ $ & $\ F_10\ $ & $\ F_22\ $ \\
 \hline
 & $+$ & $-$ & $+$ & $-$ & $-$ & $+$ \\
\end{tabular}
\caption{The $\hat S_y$ symmetry or antisymmetry indices for the representations
having $\mu=0$ and~$2$.
}
\label{tab:Sysym}
\end{table*}

%------------------------------------------------------------------------------
\subsection{Properties of the wave functions in the presence of tetrahedral-symmetry}
\label{sec:Campl}

The expansion coefficients in Eq.~(\ref{eq:lambdawf}) can be represented as
\begin{align}
       C^{\pi}_{IK,\lambda\mu\beta}
       =
       \langle I^{\pi}K|I^{\pi}\lambda\mu\beta\rangle ,
                                                   \label{eq:Ccoef}
\end{align}
and satisfy the orthonormality condition
\begin{align}
       \sum_K C^{\pi *}_{IK,\lambda'\mu'\beta'}C^{\pi}_{IK,\lambda\mu\beta}
       =
       \delta_{\lambda\lambda'} \delta_{\mu\mu'} \delta_{\beta\beta'}\,.
                                                   \label{eq:Cnorth}
\end{align}
They can be constructed in various ways.
As an example, one can obtain the coefficients in Eq.~(\ref{eq:Ccoef})
according to the group theory considerations and the angular-momentum coupling.
Consider the $A_1$ representation:
Its lowest possible $I \ne 0$ state is $I^{\pi}=3^-$.
Considering the value of the $z$-doublex $\mu=0$
and the $\hat S_y$ symmetry in the previous section,
one obtains $C^-_{3K=\pm 2,A_10}=1/\sqrt{2}$ and zero otherwise
(the additional quantum number $\beta$ is unnecessary in this case since as it can be seen from  Table VI in Ref.~\cite{TSD13} there is only one state $3^-$ in the $A_1$ representation).
Then the coupling $(3^-)_{A_1}\otimes(3^-)_{A_1}=(4^+)_{A_1}\oplus(6^+)_{A_1}$
gives the coefficients for the $I^{\pi}=4^+$ and $6^+$ states.
This process can be continued to obtain all
the expansion coefficients for the $A_1$ representation:
Those of $A_2$ are easily obtained
because $A_2$ is the parity conjugate to $A_1$.
Others can be obtained by coupling the $A_1$ states and
the lowest possible $I^{\pi}$ state of other representations
because $A_1 \otimes \lambda=\lambda$.
Although all the expansion coefficients can be obtained
in principle in this way,
it is tedious to perform such calculations for high-spin states.

An alternative way of obtaining these coefficients is via numerical diagonalization of
the projection operator onto the representation~$\lambda$,
\begin{align}
       \hat P^{[\lambda]}
       \equiv
       \frac{f_\lambda}{N_g} \sum_{g\in G}
            \chi^{[\lambda]*}(g) \hat D (g) \,,
                                                    \label{eq:lamproj}
\end{align}
within the space of $\{ |I^{\pi}K\rangle;K=-I,\cdots,I \}$.
Here $N_g$ is the number of group elements,
$\chi^{[\lambda]}(g)\equiv \sum_{\mu}D^{[\lambda]}_{\mu\mu}(g)$
is the character of $g \in G$ for the representation~$\lambda$, cf.~Ref.~~\cite{Ham62},
and $\hat D(g)$ is a group representation, cf.~Eq.~(\ref{eq:tope}).
This is a general way to construct basis states for an arbitrary
representation of the point group.
With the help of projection operator in Eq.~(\ref{eq:lamproj})
the occurrence number $n^{\lambda}_{I\pi}$ in Eq.~(\ref{eq:lambdawf}) can be calculated as
\begin{align}
       n^{\lambda}_{I\pi}=\frac{1}{f_{\lambda}}
      \sum_K \langle I^{\pi}K| \hat P^{[\lambda]}|I^{\pi}K\rangle \,,
                                                  \label{eq:nlamI}
\end{align}
from which $\sum_{\lambda}f_{\lambda}\,n^{\lambda}_{I\pi}=2I+1$ follows
because $\sum_{\lambda} \hat P^{[\lambda]}=1$.

Below we will explicitly construct
the tetrahedral-symmetric basis states $|I^\pi \lambda\mu\beta\rangle$
for the core plus one-particle systems with $\lambda=E_{1/2}$ of $T^D_d$
by coupling the even systems belonging to
$A_1$ irreducible representation of $T_d$
to the lowest spin $E_{1/2}$ system with $j^{\pi}=\frac{1}{2}^+$.
In the same way, those with $E_{5/2}$ and $G_{3/2}$
are constructed by coupling $A_2$ and $E$, respectively,
to the lowest spin $E_{1/2}$ system with $j^{\pi}=\frac{1}{2}^+$.
The underlying coupling properties follow from the direct-product properties,
$A_1 \otimes E_{1/2}=E_{1/2}$, $A_2 \otimes E_{1/2}=E_{5/2}$
and $E \otimes E_{1/2}=G_{3/2}$, respectively.
With this construction the Coriolis-coupling matrix elements
in Eq.~(\ref{eq:Hcor}) can be diagonalized analytically
for the $E_{1/2}$ and $E_{5/2}$ representations
(see Appendices~\ref{sec:ApPropwf} and~\ref{sec:ApEvC} for details).
In this way one obtains the rotational energy expressions
for the $E_{1/2}$ and $E_{5/2}$ representations.
For the case of $\lambda=G_{3/2}$ we are not able to obtain
energy expression analytically with this construction;
only an approximate expression is obtained.
In the general case of $G_{3/2}$ representation the expansion coefficients
obtained numerically from the projection operator
in Eq.~(\ref{eq:lamproj}) were employed.

Without loss of generality, we choose the same phase convention
for the coefficients in Eq.~(\ref{eq:Ccoef}) as that
of the angular-momentum state $|I^\pi K\rangle$, see e.g. Ref.~\cite{BM75};
i.e., the action of the simplex operator $\hat S_y$
and of the time-reversal operators $\hat{\cal T}$
on the wave function in Eq.~(\ref{eq:lambdawf}) are the same:
\begin{align}
       \hat S_y|I^{\pi}\lambda\mu\beta\rangle
       =
       \hat{\cal T}|I^{\pi}\lambda\mu\beta\rangle \,,
                                                     \label{eq:Trev}
\end{align}
which leads to
\begin{align}
       \pi C^{\pi}_{IK,\lambda\mu\beta}
       =
       C^{\pi *}_{IK,\lambda\mu\beta} \,,
                                                     \label{eq:Ccomp}
\end{align}
namely, the expansion coefficients are real for $\pi=+$
and pure imaginary for $\pi=-$.
The same phase convention is employed for the single-particle states.

%------------------------------------------------------------------------------
\subsection{Coriolis coupling for the \boldmath $E_{1/2}$ representation}
\label{sec:CoriE12}

The $E_{1/2}$ representation is two dimensional
with $z$-doublex $\mu=\pm 1/2$.
Because of the $y$-simplex symmetry in Eq.~(\ref{eq:Sybasis}),
we choose
\begin{align}
      |E_{1/2}-1/2\rangle=\hat S_y|E_{1/2}1/2\rangle
                                                  \label{eq:E12Sy}
\end{align}
for both the collective and single-particle wave-functions.
Then, for $(\lambda\mu)=(E_{1/2}1/2)$,
taking into account Eq.~(\ref{eq:JcorjI}),
the strong-coupling wave function can be written as
\begin{align}
      | \Psi^{\lambda}_{I^\pi \beta} \rangle
      &=
      \frac{1}{\sqrt{2}}
      \left[1+\hat \Pi \hat \pi\, e^{i\pi (\hat \jmath_2-\hat I_2)} \right]
      |I^\pi \lambda \mu \beta \rangle
      | \phi_{\lambda \mu} \rangle
                                                  \nonumber\\
      &=
      \frac{1}{\sqrt{2}}
      \left[|I^\pi \lambda \mu \beta \rangle | \phi_{\lambda \mu} \rangle
      +
      \hat S_y|I^\pi \lambda \mu \beta \rangle
      \hat \pi e^{i\pi \hat \jmath_2}| \phi_{\lambda \mu} \rangle \right],
                                                  \label{eq:E15swf}
\end{align}
and the Coriolis-coupling matrix element is given by
\begin{align}
      &\langle \Psi^{\lambda}_{I^\pi \beta'}
      |2\mbold{\hat I}\cdot\mbold{\hat \jmath}
      | \Psi^{\lambda}_{I^\pi \beta} \rangle
      =
      \langle \Psi^{\lambda}_{I^\pi \beta'} |
      \hat I_+ \hat \jmath_- +\hat I_- \hat \jmath_+
      +2\hat I_3 \hat \jmath_3
      | \Psi^{\lambda}_{I^\pi \beta} \rangle
                                                   \nonumber\\
      &
      =\langle I^\pi \lambda \mu \beta' | \hat I_- \hat S_y
      | I^\pi \lambda \mu \beta \rangle
      \langle \phi_{\lambda \mu} | \hat \jmath_+ \hat \pi e^{i\pi \hat \jmath_2}
      | \phi_{\lambda\mu} \rangle
      +2\langle I^\pi \lambda \mu \beta' | \hat I_3
      | I^\pi \lambda \mu \beta \rangle
      \langle \phi_{\lambda \mu} |\hat \jmath_3 | \phi_{\lambda \mu} \rangle ,
                                                   \label{eq:E15CoriC}
\end{align}
where the relations $\hat {S_y}^{\dagger} \hat I_3 \hat S_y=-\hat I_3$ and
the similar one for $\hat \jmath_3$ have been used.
Note that $\hat I_{+}$  ($\hat I_{-}$) decreases (increases) $\mu$
by one unit (mod 4).
It can be seen that the wave function in Eq.~(\ref{eq:E15swf})
has essentially the same form as the ${\cal R}$-invariant wave function
for the axially symmetric rotational band in Sec.~4-2c of Ref.~\cite{BM75}
(in fact, the signature operation should be
replaced by the simplex operation).

As already mentioned we construct
a specific $E_{1/2}$ collective basis wave-function with $\mu=1/2$
by coupling the $A_1$ basis states with that of
the smallest spin positive-parity state of $E_{1/2}$, $|\mhalf^+\mhalf\rangle$.
In fact, it is possible because $A_1\otimes E_{1/2}=E_{1/2}$,
and it is enough because
$n^{E_{1/2}}_{I\pi}=n^{A_1}_{(I-1/2)\pi}+n^{A_1}_{(I+1/2)\pi}$;
i.e., all the basis states are generated in this way
(in obtaining these relations the information contained in Tables VI and VIII
of Ref.~\cite{TSD13} has been used). Thus,
\begin{align}
      {\cal N}^{I\pi}_{\lambda \mu \alpha}| I^\pi E_{1/2}1/2 \alpha \rangle
      & \equiv
      \Bigl[ | k^{\pi} A_1 0 \gamma \rangle
      \otimes
      | \mhalf^+ \mhalf \rangle \Bigr]_I,
                                                \label{eq:E12Dbasis}
\end{align}
where ${\cal N}^{I\pi}_{\lambda \mu \alpha}$ is normalization constant
and $\alpha=(k\gamma)$ with $k=I\pm \mhalf$ and $\gamma$ denotes the additional
quantum number for the basis states of $A_1$.
Although we are not able to prove it generally, we have confirmed that
operators $\hat I_3$ and of $\hat I_- \hat S_y$ appearing
in the Coriolis coupling in Eq.~(\ref{eq:E15CoriC})
are diagonal within these specific basis states $(\beta\rightarrow\alpha)$.
If the numerically calculated basis states by diagonalizing
the projection operator in Eq.~(\ref{eq:lamproj}) are employed,
the matrix elements of $\hat I_3$ are not diagonal
and it turns out that the eigenvalues are $(I+1)/3$ and $-I/3$,
corresponding to Eq.~(\ref{eq:defglamI}).

The diagonal matrix-elements in Eq.~(\ref{eq:E15CoriC}) with
the basis state in Eq.~(\ref{eq:E12Dbasis}) can be evaluated
by using the identities of the expansion coefficients of $A_1$.
The details are presented in Appendices~\ref{sec:ApPropwf} and~\ref{sec:ApEvC}, whereas the result of interest reads:
\begin{align}
 &2 \langle I^\pi E_{1/2}1/2 \alpha'| \hat I_3
 | I^\pi E_{1/2}1/2 \alpha \rangle
 =-\langle I^\pi E_{1/2}1/2 \alpha'| \hat I_- \hat S_y
  | I^\pi E_{1/2}1/2 \alpha \rangle
 \nonumber \\
 &\quad = -g_{A_1}(I)\,\delta_{\alpha'\alpha},
\label{eq:CorIE12r}
\end{align}
where the function $g_{A_1}(I)$ is defined by the following generic expression $g_{\lambda}(I)$
with $\lambda=A_1$;
\begin{align}
 g_{\lambda}(I)\equiv\frac{2}{3}\times
 \biggl\{\begin{matrix}
 -(I+1), & I=I_{\lambda}+\mhalf, \cr
   I,    & I=I_{\lambda}-\mhalf, \end{matrix}\biggr.
\label{eq:defglamI}
\end{align}
with $I_{\lambda}$ representing the allowed values of angular-momentum
within  $\lambda$-representation.
Then the energy spectrum is given by
one parameter, here denoted as $a^{E_{1/2}}$;
\begin{align}
       E_{E_{1/2}}(I)=\frac{1}{2{\cal J}}\bigr[ I(I+1)
       +
       a^{E_{1/2}} g_{A_1}(I) \bigr],
                                                  \label{eq:EdecE12}
\end{align}
defined by
\begin{align}
       a^{E_{1/2}}
       \equiv
       \langle \phi_{E_{1/2}1/2} | -\hat \jmath_+ \hat \pi e^{i\pi \hat \jmath_2}
      +\hat \jmath_3 | \phi_{E_{1/2}1/2} \rangle \,.
                                                  \label{eq:aE12}
\end{align}
Consequently, the spectrum splits into two parabolas according to
$I=I_{A_1}\pm \mhalf$, and the amount of splitting is determined
by the generalized decoupling parameter, $a^{E_{1/2}}$.
This result is structurally similar to the one valid in the case of the axial symmetry, cf.~Sec.~\ref{sec:axsym}, Eqs.~(\ref{eq:Edecax}) and~(\ref{eq:aaxK12}).

%------------------------------------------------------------------------------
\subsection{Coriolis coupling for the \boldmath $E_{5/2}$ representation}
\label{sec:CoriE52}

The $E_{5/2}$ representation is parity-conjugate of $E_{1/2}$ and
has $z$-doublex $\mu=\mp 3/2$.
The basis state can be constructed by coupling the $A_2$ basis states
with $|\mhalf^+\mhalf\rangle$ because $A_2 \otimes E_{1/2}=E_{5/2}$
(or equivalently, one can construct it by coupling the $A_1$ basis states
with the smallest spin-parity state of $E_{5/2}$,
$|\mhalf^-\mhalf\rangle$, because $A_1 \otimes E_{5/2}=E_{5/2}$).
Again, this gives all the basis states
because $n^{E_{5/2}}_{I\pi}=n^{A_2}_{(I-1/2)\pi}+n^{A_2}_{(I+1/2)\pi}$.
Note that the corresponding $z$-doublex exponent of the resulting wave function satisfies $\mu=1/2+2\equiv -3/2$ (mod~4),
and consequently,
\begin{align}
 | I^\pi E_{5/2}-3/2 \alpha \rangle
 \propto \Bigl[ | k^{\pi} A_2 2 \gamma \rangle
 \otimes | \mhalf^+ \mhalf \rangle \Bigr]_I,
                                                   \label{eq:E52Dbasis}
\end{align}
whereas the simplex-conjugate state is defined by
\begin{align}
 |E_{5/2}3/2\rangle=\hat S_y|E_{5/2}-3/2\rangle.
\label{eq:E52Sy}
\end{align}
One shows that the structure of the wave functions in the $E_{5/2}$ representation is analogous to the one in Eq.~(\ref{eq:E15swf})
with $(\lambda\mu)=(E_{5/2}-3/2)$.
Here, similar calculations can be performed as in the case of $E_{1/2}$,
with the only difference  that $A_2$ has opposite $\hat S_y$ symmetry
to $A_1$ as shown in Table~\ref{tab:Sysym}
(see Appendices~\ref{sec:ApPropwf} and~\ref{sec:ApEvC} for details).
The matrix elements for the $E_{5/2}$ representation are then given by
\begin{align}
 &2 \langle I^\pi E_{5/2}-3/2 \alpha'| \hat I_3
 | I^\pi E_{5/2}-3/2 \alpha \rangle
  = \langle I^\pi E_{5/2}-3/2 \alpha'| \hat I_- \hat S_y
  | I^\pi E_{5/2}-3/2 \alpha \rangle
 \nonumber \\
 &\quad =-g_{A_2}(I)\,\delta_{\alpha'\alpha},
                                                    \label{eq:CorIE52r}
\end{align}
where $g_{A_2}(I)$ is defined by Eq.~(\ref{eq:defglamI})
with $\lambda=A_2$. The corresponding spectrum is given by
\begin{align}
       E_{E_{5/2}}(I)=\frac{1}{2{\cal J}}
       \bigr[ I(I+1) + a^{E_{5/2}} g_{A_2}(I) \bigr],
                                                    \label{eq:EdecE52}
\end{align}
where the generalized decoupling parameter is defined as
\begin{align}
       a^{E_{5/2}}
       \equiv
       \langle \phi_{E_{5/2}-3/2} |
       \hat \jmath_+ \hat \pi e^{i\pi \hat \jmath_2}
       +
       \hat \jmath_3 | \phi_{E_{5/2}-3/2} \rangle .
                                                    \label{eq:aE52}
\end{align}
This result is similar to the case with the axial symmetry, cf.~Eqs.~(\ref{eq:Edecax}) and~(\ref{eq:aaxK12}) in Sec.~\ref{sec:axsym}.

%------------------------------------------------------------------------------
\subsection{Coriolis coupling for the \boldmath $G_{3/2}$ representation}
\label{sec:CoriG32}

It can be demonstrated that the $T_d^D$ covariant wave function for the $G_{3/2}$ representation
has four components with the $z$-doublex exponent $\mu=\pm 1/2,\mp 3/2$:
\begin{align}
 | \Psi^{G_{3/2}}_{I^\pi \beta} \rangle
 &=\frac{1}{\sqrt{4}} \sum_{\mu=\pm 1/2, \mp 3/2}
 |I^\pi G_{3/2} \mu \beta \rangle | \phi_{G_{3/2} \mu} \rangle
 \nonumber \\
 &= \frac{1}{\sqrt{4}}\sum_{\mu=1/2,-3/2}
  \left[1+\hat \Pi \hat \pi e^{i\pi (\hat j_2-\hat I_2)} \right]
 |I^\pi G_{3/2} \mu \beta \rangle | \phi_{G_{3/2} \mu} \rangle
 \nonumber \\
 &=  \frac{1}{\sqrt{4}} \sum_{\mu=1/2,-3/2}
  \left[|I^\pi G_{3/2} \mu \beta \rangle | \phi_{G_{3/2} \mu} \rangle
   +\hat S_y|I^\pi G_{3/2} \mu \beta \rangle
   \hat \pi e^{i\pi \hat j_2}| \phi_{G_{3/2} \mu} \rangle \right],
\end{align}
where the simplex-conjugate states are defined
as the case of $E_{1/2}$-representation in Eq.~(\ref{eq:E12Sy})
or $E_{5/2}$-representation of Eq.~(\ref{eq:E52Sy}), i.e.:
\begin{align}
      |G_{3/2}-\mu\rangle=\hat S_y|G_{3/2}\mu\rangle,
      \quad
      \mu=1/2,-3/2.
                                                      \label{eq:G32Sy}
\end{align}
The sought Coriolis-coupling matrix elements are given by
\begin{align}
 &\langle \Psi^{G_{3/2}}_{I^\pi \beta'} |
 2\mbold{\hat I \cdot \hat \jmath} | \Psi^{G_{3/2}}_{I^\pi \beta} \rangle
 \nonumber \\
 & =\frac{1}{2} \left[
 \langle I G_{3/2} 1/2 \beta'|\hat I_- \hat S_y |I G_{3/2} 1/2 \beta \rangle
 \langle \phi_{G_{3/2} 1/2} | \hat \jmath_+ \hat \pi e^{i\pi \hat \jmath_2}
 |  \phi_{G_{3/2} 1/2} \rangle
 \right. \nonumber \\
 &\qquad+
 \langle I G_{3/2} -3/2 \beta'|\hat I_- \hat S_y |I G_{3/2} -3/2 \beta \rangle
 \langle \phi_{G_{3/2} -3/2} | \hat \jmath_+ \hat \pi e^{i\pi \hat \jmath_2}
 | \phi_{G_{3/2} -3/2} \rangle
 \nonumber \\
 &\qquad+
 \langle I G_{3/2} -3/2 \beta'|\hat I_+ \hat S_y |I G_{3/2} 1/2 \beta \rangle
 \langle \phi_{G_{3/2} -3/2} | \hat \jmath_- \hat \pi e^{i\pi \hat \jmath_2} |
  \phi_{G_{3/2} 1/2} \rangle
 \nonumber \\
 &\qquad+
 \langle I G_{3/2} 1/2 \beta'|\hat I_+ \hat S_y |I G_{3/2} -3/2 \beta \rangle
 \langle \phi_{G_{3/2} 1/2} | \hat \jmath_- \hat \pi e^{i\pi \hat \jmath_2} |
  \phi_{G_{3/2}-3/2} \rangle
 \nonumber \\
 &\qquad+
 2\langle I G_{3/2} 1/2 \beta'|\hat I_3 | I G_{3/2} 1/2 \beta \rangle
 \langle \phi_{G_{3/2} 1/2} | \hat \jmath_3 |  \phi_{G_{3/2} 1/2} \rangle
 \nonumber \\
 &\qquad+
 \left.2\langle I G_{3/2} -3/2 \beta'|\hat I_3 | I G_{3/2} -3/2 \beta \rangle
 \langle \phi_{G_{3/2} -3/2} | \hat \jmath_3 |  \phi_{G_{3/2} -3/2} \rangle
 \right].
\label{eq:G32CoriC}
\end{align}
Note that a new type of matrix elements of
the operator $\hat I_+\hat S_y$ between
$\mu=1/2$ and $-3/2$ states appear compared with
the cases of $E_{1/2}$ and/or $E_{5/2}$ in Eq.~(\ref{eq:E15CoriC}).

Thus, for $G_{3/2}$-representation, there are six types of collective matrix elements
in the Coriolis coupling in Eq.~(\ref{eq:G32CoriC}).
However, we have found by numerical calculations
that only two of them are independent.
In order to see the relations between these matrix elements
it is necessary to fix
the relative phase between the $\mu=1/2$ and $-3/2$ states.
For this purpose we construct the $\mu=-3/2$ state from the $\mu=1/2$ state using a specially constructed shift operator satisfying
\begin{align}
 | G_{3/2} -3/2\rangle = \hat X_+| G_{3/2} 1/2\rangle .
                                                       \label{eq:G32X}
\end{align}
It will be shown in the Appendix~\ref{sec:ApBsG32} that
\begin{align}
      \hat X_+\equiv i\sqrt{\frac{2}{3}}\,\Bigl(\hat S^{}_4+ \hat S^{\dagger}_4
      +\frac{1}{\sqrt{2}}\Bigr),
                                                      \label{eq:G32Xop}
\end{align}
where the operator $\hat S_4$ is a class $S_4$ group element
of $T_d$ (and of $T_d^D$) and is defined
in the Appendix~\ref{sec:ApPropwf} (Eq.~(\ref{eq:S4xp}));
see Appendix~\ref{sec:ApBsG32} for details.
We have constructed the collective basis states with $\mu=1/2$ by
numerically diagonalizing the projection operator in Eq.~(\ref{eq:lamproj}),
and other states are obtained by Eqs.~(\ref{eq:G32X}) and~(\ref{eq:G32Sy}).
With these basis states, we obtain the following
relations between the six matrix elements in Eq.~(\ref{eq:G32CoriC}):
\begin{align}
 &\langle I G_{3/2} 1/2 \beta'|\hat I_3 | I G_{3/2} 1/2 \beta \rangle
 +\langle I G_{3/2} -3/2 \beta'|\hat I_3 | I G_{3/2} -3/2 \beta \rangle
 \nonumber \\
 &=
 \frac{1}{2}\Bigl[
 -\langle I G_{3/2} 1/2 \beta'|\hat I_- \hat S_y |I G_{3/2} 1/2 \beta \rangle
 +\langle I G_{3/2} -3/2 \beta'|\hat I_- \hat S_y |I G_{3/2} -3/2 \beta \rangle
 \Bigr]
 \nonumber \\
 &\equiv \Lambda^1_{\beta'\beta},
\label{eq:G32relME1}
\end{align}
and
\begin{align}
 &\langle I G_{3/2} 1/2 \beta'|\hat I_3 | I G_{3/2} 1/2 \beta \rangle
 -\langle I G_{3/2} -3/2 \beta'|\hat I_3 | I G_{3/2} -3/2 \beta \rangle
 \nonumber \\
 &=
 \langle I G_{3/2} 1/2 \beta'|\hat I_- \hat S_y |I G_{3/2} 1/2 \beta \rangle
 +\langle I G_{3/2} -3/2 \beta'|\hat I_- \hat S_y |I G_{3/2} -3/2 \beta \rangle
 \nonumber \\
 &=\frac{2}{\sqrt{3}}
 \langle I G_{3/2} -3/2 \beta'|\hat I_+ \hat S_y |I G_{3/2} 1/2 \beta \rangle
 \equiv \Lambda^2_{\beta'\beta},
\label{eq:G32relME2}
\end{align}
together with the following identity:
\begin{align}
 \langle I G_{3/2} -3/2 \beta'|\hat I_+ \hat S^{}_y |I G_{3/2} 1/2 \beta \rangle
 &=\langle I G_{3/2} 1/2 \beta| \hat S^{\dagger}_y \hat I_-|
 I G_{3/2} -3/2 \beta'\rangle^*
 \nonumber \\
 &=\langle I G_{3/2} 1/2 \beta|\hat I_+ \hat S^{}_y |
I G_{3/2} -3/2 \beta'\rangle^*.
\label{eq:G32relME3}
\end{align}
The latter follows because
$(\hat S_y)^2 |I^\pi \lambda\mu\beta\rangle=-|I^\pi \lambda\mu\beta\rangle$
for half-integer values of $I$ and
$\hat S^{\dagger}_y \hat I_-\hat S^{}_y=-\hat I_+$.
With the phase convention in Eq.~(\ref{eq:Trev}) the matrix elements
of $I_3$ are real and the corresponding matrix is symmetric; similar can be said about all the six collective
matrix elements implied by the above relations.

Using Eqs.~(\ref{eq:G32relME1})$-$(\ref{eq:G32relME3})
all the six collective matrix elements in Eq.~(\ref{eq:G32CoriC})
can be expressed in terms of the two matrices,
$\Lambda^1_{\beta'\beta}$ and $\Lambda^2_{\beta'\beta}$,
which are also real and symmetric,
and the Coriolis coupling can be written as
\begin{align}
 &\langle \Psi^{G_{3/2}}_{I^\pi \beta'} |
 2\mbold{\hat I \cdot \hat \jmath} | \Psi^{G_{3/2}}_{I^\pi \beta} \rangle
 =a^{G_{3/2}}_{1} \,\Lambda^1_{\beta'\beta}
 +a^{G_{3/2}}_{2} \,\Lambda^2_{\beta'\beta},
 \qquad
\label{eq:G32CorCa}
\end{align}
where two generalized decoupling parameters which appear in this case are defined by
\begin{align}
 a^{G_{3/2}}_{1}&=\frac{1}{2}\bigl[
 \langle \phi_{G_{3/2} 1/2}
 | -\hat \jmath_+ \hat \pi e^{i\pi \hat \jmath_2}+\hat \jmath_3 |
  \phi_{G_{3/2} 1/2} \rangle
 +\langle \phi_{G_{3/2} -3/2}
 | \hat \jmath_+ \hat \pi e^{i\pi \hat \jmath_2}+\hat \jmath_3 |
  \phi_{G_{3/2} -3/2} \rangle\bigr],
                                                \label{eq:a1G32} \\
 a^{G_{3/2}}_{2}&= \frac{1}{4}\bigl[
 \langle \phi_{G_{3/2} 1/2}
 | \hat \jmath_+ \hat \pi e^{i\pi \hat \jmath_2}+2\hat \jmath_3 |
  \phi_{G_{3/2} 1/2} \rangle
 +\langle \phi_{G_{3/2} -3/2}
 | \hat \jmath_+ \hat \pi e^{i\pi \hat \jmath_2} - 2\hat \jmath_3 |
  \phi_{G_{3/2} -3/2} \rangle\bigr]
                                                \nonumber \\
 &\qquad + \frac{\sqrt{3}}{2}\langle \phi_{G_{3/2} -3/2} |
 \hat \jmath_- \hat \pi e^{i\pi \hat \jmath_2}
 |  \phi_{G_{3/2} 1/2} \rangle\,.
                                                \label{eq:a2G32}
\end{align}
We have used the fact that all the intrinsic matrix elements are real
within the adopted phase convention. Expressions $| \phi_{G_{3/2} -3/2} \rangle$
in Eqs.~(\ref{eq:a1G32}) and~(\ref{eq:a2G32}) should be calculated analogously as
\begin{align}
 &| \phi_{G_{3/2} -3/2} \rangle=\hat x_+ |  \phi_{G_{3/2} 1/2} \rangle,
 \\
 &\hat x_+=i \sqrt{\frac{2}{3}}\,
 \Bigl(\hat s_4+\hat s_4^\dagger -\frac{1}{\sqrt{2}}\Bigr),
 \quad
 \hat s_4=\hat \pi\, e^{i\frac{\pi}{4}\hat \jmath_3}
  e^{i\frac{\pi}{2}\hat \jmath_2} e^{-i\frac{\pi}{4}\hat \jmath_3},
\label{eq:G32x}
\end{align}
from which Eqs.~(\ref{eq:lrepe}) and~(\ref{eq:lrepi}) follow
with the common representation matrix, $D^{[G_{3/2}]}_{\mu\mu'}(g)$.
Therefore, the Coriolis coupling in the $G_{3/2}$ representation
cannot be calculated analytically in contrast to the cases
of $E_{1/2}$ and $E_{5/2}$, and the coupling Hamiltonian
should be diagonalized numerically to obtain the spectra.

However, it is interesting to note that an approximate expression for the energy levels
can be obtained for the particular case of
$a^{G_{3/2}}_{2} \approx 0$
in Eq.~(\ref{eq:G32CorCa});
\begin{align}
       E_{G_{3/2}}(I)
       \approx
       \frac{1}{2{\cal J}}
       \bigr[ I(I+1) + a^{G_{3/2}}_{1} g_{E}(I) \bigr]
       \;\; {\rm for}\;\;
       a^{G_{3/2}}_{2} \approx 0,
                                                      \label{eq:EdecG32}
\end{align}
where $g_{E}(I)$ is defined by Eq.~(\ref{eq:defglamI}) with $\lambda=E$.
In this case the spectrum splits into two sequences
like in the cases of the $E_{1/2}$ and $E_{5/2}$ representations.
This can be confirmed by taking a specific basis states of
the $G_{3/2}$ representation,
which is constructed by coupling the $E$ basis states
with $|\mhalf^+\mhalf\rangle$ because $E \otimes E_{1/2}=G_{3/2}$.
This coupling gives all needed basis states
because $n^{G_{3/2}}_{I\pi}=n^{E}_{(I-1/2)\pi}+n^{E}_{(I+1/2)\pi}$.
Thus, the basis states with $\mu=1/2$ and $\mu=-3/2$ are constructed by
\begin{equation}
\left.
\begin{array}{lc}
 | I^\pi G_{3/2}1/2 \alpha \rangle
 &\propto \Bigl[ | k^{\pi} E 0 \gamma \rangle
 \otimes | \mhalf^+ \mhalf \rangle \Bigr]_I,
 \cr
 | I^\pi G_{3/2}-3/2 \alpha \rangle
 &\propto \Bigl[ | k^{\pi} E 2 \gamma \rangle
 \otimes | \mhalf^+ \mhalf \rangle \Bigr]_I,
\end{array}\right.
\label{eq:G32mbasis}
\end{equation}
where $\alpha=(k\gamma)$ with $k=I\pm \mhalf$.
The simplex conjugate states are defined by Eq.~(\ref{eq:G32Sy}).
As in the cases of $E_{1/2}$ and of $E_{5/2}$,
the matrix relation in Eq.~(\ref{eq:G32relME1}) can be confirmed
for this specific basis state ($\beta\rightarrow\alpha$)
(see Appendices~\ref{sec:ApPropwf} and~\ref{sec:ApEvC} for details),
and one finds that the coupling matrix $(\Lambda^1_{\alpha'\alpha})$
is diagonalized in these specific basis states; i.e.,
\begin{align}
 \Lambda^1_{\alpha'\alpha}=-g_{E}(I)\,\delta_{\alpha'\alpha}.
\label{eq:CorIG32r}
\end{align}
In this way the validity of Eq.~(\ref{eq:EdecG32}) can be demonstrated.
Note that even with these specific basis states another matrix
$\Lambda^2_{\alpha'\alpha}$ in Eq.~(\ref{eq:G32relME2}),
which connects the states of the $z$-doublex $\mu=1/2$ and $-3/2$,
is not diagonal
and the numerical diagonalization is necessary when $a^{G_{3/2}}_{2} \ne 0$.

%------------------------------------------------------------------------------
%------------------------------------------------------------------------------
\section{Comparison with microscopic projection calculations}
\label{sec:results}

In the present work we aim at an illustration of the approach discussed so far within its asymptotic limit in terms of the strong-coupling. We will arbitrarily select an excessively large tetrahedral deformation to assure the applicability of this starting point assumption. This will allow us to examine various mathematical details of the modeling introduced here in the possibly simplest realization of the strong coupling.
More precisely, tetrahedral deformation
of $\alpha_{32}=\alpha_{3-2}=0.40$ will be employed,
with which an ideal rotational spectrum could be obtained~\cite{TSD13}.

In the present work, we have performed
the angular-momentum and parity projection calculations
for the tetrahedral-deformed
core plus one-particle systems in $^{81}_{40}$Zr,
for which $N=Z=40$ corresponds
to the tetrahedral doubly-magic configurations,
see e.g. Ref.~\cite{DGM10}.
The method of the calculations is the same as in Ref.~\cite{TSD13},
i.e., we employ the Woods-Saxon mean-field and
the schematic separable-type interactions consistent with it,
which are composed of the $l=2,3,4$ multipole-multipole
interaction terms and of the $l=0,2$
multipole pairing interactions; the reader interested in particular in the determination of the  coupling constants in this case may consult
Ref.~\cite{TS12}.
Except for the treatment of pairing correlations,
there is one difference
with respect to the calculations in Ref.~\cite{TSD13}, {\em viz.}, to gain in simplicity,
the infinitesimal cranking~\cite{TS16} is not performed.

The static pairing correlations in the mean field are neglected
for simplicity, i.e., we assume that
the pairing gaps for both neutrons and protons vanish. Indeed, microscopic calculations indicate that for doubly-magic tetrahedral-symmetry nuclear configurations the presence of the large gaps reduces pairing correlations considerably.

For the calculation of the decoupling parameters introduced in this work,
we use the single-particle states $|\phi_{\lambda\mu}\rangle$
obtained by the same Woods-Saxon potential
as in the angular-momentum and parity projection calculations cited earlier.
In order to compare the calculated spectra obtained within the present model
with those of the microscopic projection calculations,
one has to take appropriate values of
the moment of inertia ${\cal J}$ in Eq.~(\ref{eq:Hcor}),
which is an input parameter in the present formulation.
For this purpose, we calculate the following quantity,
\begin{align}
      {\mit\Delta}E
      =
      {\bar E}_{I_{\rm f}}-{\bar E}_{I_{\rm i}},
      \quad
      {\bar E}_I
      \equiv
      \frac{\sum_{\pi \beta} E_{I\pi\beta}}{\sum_{\pi \beta} 1},
                                                        \label{eqn.62}
\end{align}
in both the microscopic projection and the Coriolis-coupling model calculations,
and the moment of inertia was determined so that the two results coincide.
Presently the values $I_{\rm i}=1/2$ and $I_{\rm f}=25/2$ are used
for $E_{1/2}$ and $E_{5/2}$,
and $I_{\rm i}=3/2$ and $I_{\rm f}=25/2$ for $G_{3/2}$.

To generate the spectra of the core plus one-particle systems,
we place one neutron at a single-particle state
above the $N=40$ tetrahedral magic number.
Such single-particle states at the tetrahedral deformation
$\alpha_{32}=0.4$ are calculated to be
$G_{3/2}$, $E_{5/2}$, $G_{3/2}$, $G_{3/2}$, $E_{1/2}$, $\cdots$
in the order of energy.
It should be mentioned that the final spectra obtained by selecting
one of the two double-degenerate states of
the $E_{1/2}$ or of $E_{5/2}$ orbitals,
or one among the four degenerate states of the $G_{3/2}$ orbital do not depend on which one of the degenerate orbitals has been actually selected;
we have confirmed this by the microscopic projection calculations.

%------------------------------------------------------------------------------
\subsection{Results for the $\mbold{E_{1/2}}$ and $\mbold{E_{5/2}}$ cases}
\label{sec:rltE15}

\begin{figure}[!htb]
\begin{center}
\includegraphics[width=75mm]{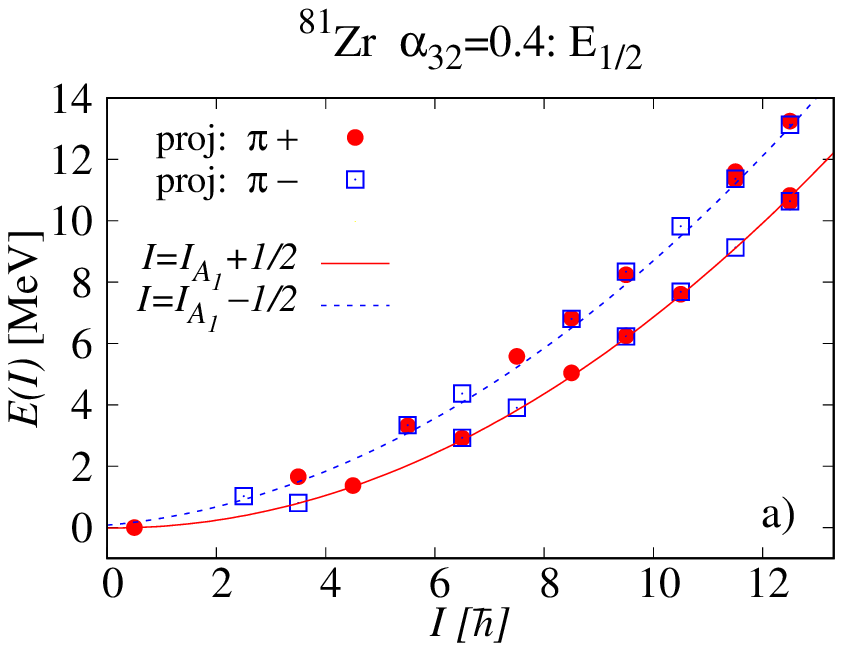}
\hspace*{-3mm}
\includegraphics[width=75mm]{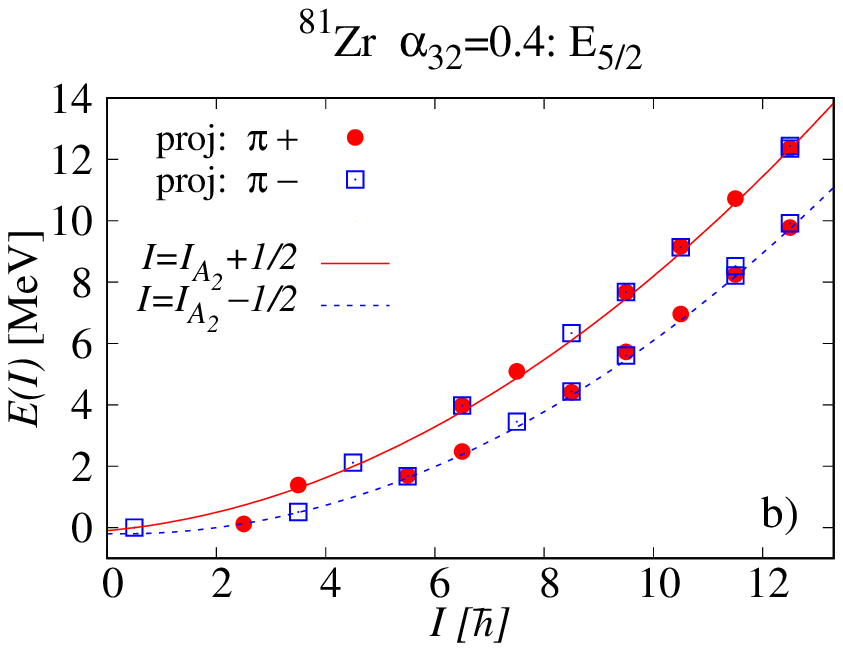}
\end{center}
\vspace*{-8mm}
\caption{(Color online)
Excitation spectra calculated
by the angular-momentum and parity projection method represented by full circles for parity~$+$ and open squares for parity~$-$,
for the core plus one-particle system $^{81}$Zr.
The left (right) panel shows the results obtained when occupying the lowest neutron $E_{1/2}$ ($E_{5/2}$) orbital
above the $N=40$ tetrahedral magic shell closure.
The solid and dotted lines are the results of the present Coriolis-coupling model
in Eq.~(\ref{eq:EdecE12}) (left panel) and
in Eq.~(\ref{eq:EdecE52}) (right panel),
where $I_{A_1}^\pi=0^+,3^-,4^+,6^{\pm},\cdots$
and $I_{A_2}^\pi=0^-,3^+,4^-,6^{\mp},\cdots$ are the allowed spin-parity
of the $A_1$ and $A_2$ representations, respectively.
}
\label{fig:E15}
\end{figure}

The results of the calculations are shown in Fig.~\ref{fig:E15} for the case of $E_{1/2}$
(left panel) and of $E_{5/2}$ (right panel),
where the results of the Coriolis-coupling model are also included
as the solid and dotted lines.
Here the lowest-energy orbital belonging to $E_{1/2}$ or to $E_{5/2}$
is occupied by the odd neutron.
In these cases, the calculated decoupling parameters
and the moment of inertia are
\begin{align}
 a^{E_{1/2}}=1.86,\quad {\cal J}=7.10\ \mbox{[$\hbar^2$/MeV]},
\label{para:E12}
\end{align}
and
\begin{align}
 a^{E_{5/2}}=-2.27,\quad {\cal J}=7.54\ \mbox{[$\hbar^2$/MeV]},
\label{para:E52}
\end{align}
for the rotational bands belonging to the irreducible representations $E_{1/2}$ and $E_{5/2}$,
respectively, corresponding to the left and right panels in Fig.~\ref{fig:E15}.

It is remarkable that the results of the microscopic projection
calculation and of the simple energy expressions obtained
by the Coriolis coupling agree to far extent
in both the $E_{1/2}$ and $E_{5/2}$ cases (up to a single adjustable constant, cf.~Eq.~(\ref{eqn.62})).
This is non-trivial because no presence of any ``rotor'' contribution is assumed
in the microscopic part of the calculations.
In fact, it was shown in Ref.~\cite{TSD13} by using
the same microscopic projection approach that the specific spin-parity
states allowed by the group theory compose one rotational band
at large tetrahedral deformation for the case of the core systems
corresponding to the $A_1$ representation of the $T_d$ group.
In the present core plus one-particle systems,
for the $E_{1/2}$ and $E_{5/2}$ representations of the $T^D_d$ group,
the expected spin-parity states appear as a result of calculation,
but the spectra still split into two parabolic-type sequences as shown in Fig.~\ref{fig:E15}.
A closer look into these two sequences reveals that one is composed
of the spin-parity states of the $A_1$ ($A_2$) representation
shifted by spin $+1/2$ and another is those shifted by spin $-1/2$
for the $E_{1/2}$ ($E_{5/2}$) case.
This is exactly the consequence of the Coriolis coupling
discussed in Sec.~\ref{sec:CoriE12} (Sec.~\ref{sec:CoriE52}).
The states with $I=I_{A_1}+1/2$ are lower in the case of $E_{1/2}$
and those with $I=I_{A_2}-1/2$ are lower in the case of $E_{5/2}$
in accordance with the sign of the decoupling parameter
in these two cases, see Eqs.~(\ref{eq:EdecE12}) and~(\ref{eq:EdecE52})
compared with Eqs.~(\ref{para:E12}) and~(\ref{para:E52}).
The energy splitting between the two sequences is
also well described by these values of the decoupling parameters.
Good agreement with the results of the microscopic calculations
suggests that the simple particle-core coupling picture is valid
for the case of tetrahedral symmetry at least asymptotically at the (large deformation) strong coupling limit.
This result is very similar to the one of the $K=1/2$ rotational bands
of axially-symmetric nuclei~\cite{BM75}.

Let us remark in passing that a similar formalism can be applied to
the even-even non-core configurations, for example,
the case where two nucleons occupy a four-fold degenerate orbital $G_{3/2}$,
by using the appropriately adapted decompositions,
in this case ${\cal A}(G_{3/2}\otimes G_{3/2})=A_1\oplus E\oplus F_2$,
where ${\cal A}$ means the anti-symmetrization.
An example of the result of the microscopic projection calculation
is shown in Fig.~4 of Ref.~\cite{TSF14} for such a case,
where the feature of splittings of the rotational bands
seems to be more complicated than the core plus one-particle systems
in the present work.

%------------------------------------------------------------------------------
\subsection{Results for the $\mbold{G_{3/2}}$ case}
\label{sec:rltG32}

\begin{figure}[!htb]
\begin{center}
\includegraphics[width=75mm]{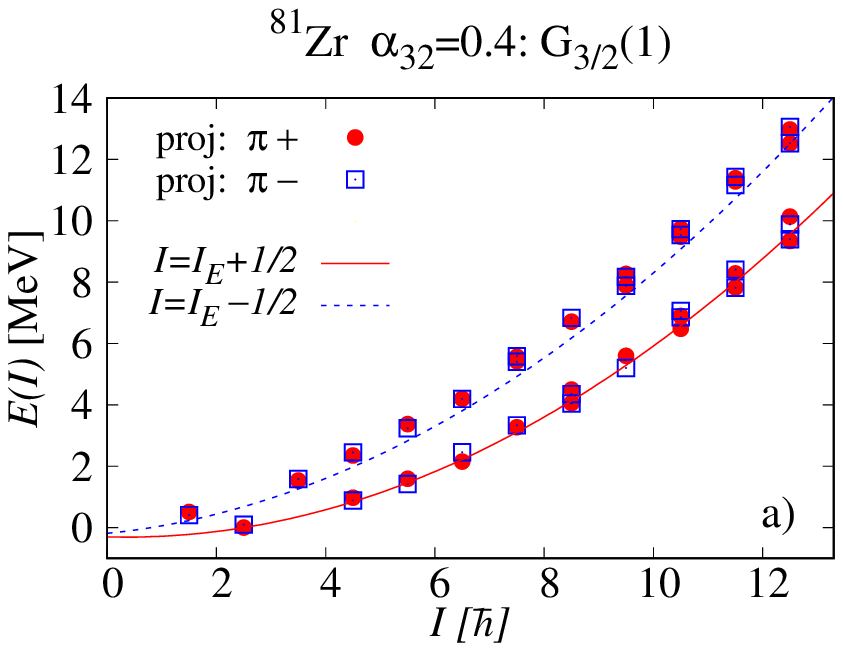}
\hspace*{-3mm}
\includegraphics[width=75mm]{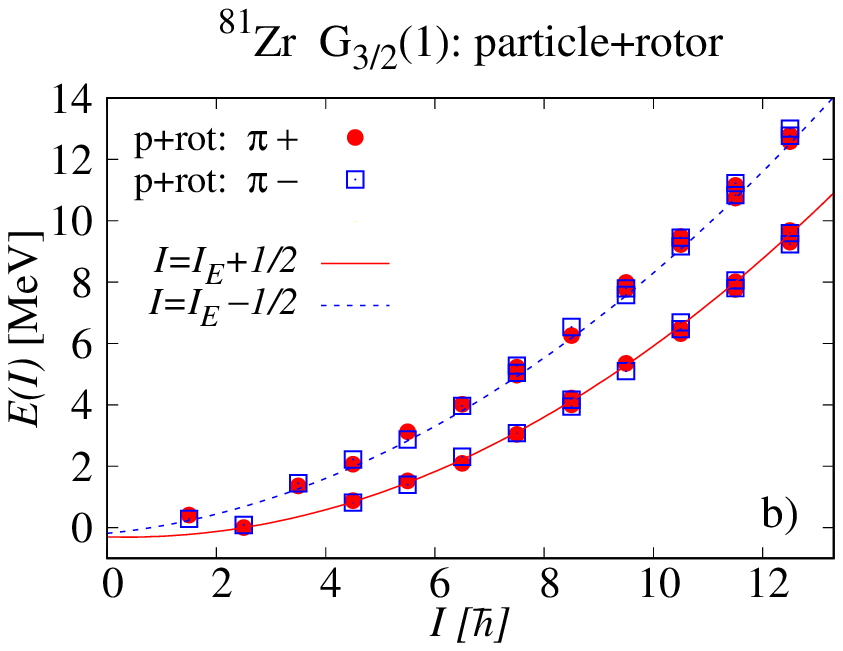}
\end{center}
\vspace*{-8mm}
\caption{(Color online)
Excitation spectra calculated
by the angular-momentum and parity projection method (left panel)
and by the Coriolis-coupling model (right panel) represented by filled circles for parity $+$ and open square for parity $-$,
for the core plus one-particle system in $^{81}$Zr,
where one neutron occupies one member-state of the lowest $G_{3/2}$ four-fold degenerate orbital
above the $N=40$ magic shell-closure.
The solid and dotted lines, shown in both the left and right panels,
are the results of the Coriolis-coupling model
with the approximation in Eq.~(\ref{eq:EdecG32}),
where $I_{E}^\pi=2^{\pm},4^{\pm},5^{\pm},6^{\pm},\cdots$
is the allowed spin-parity of the $E$ representation.
}
\label{fig:G32a}
\end{figure}

\begin{figure}[!htb]
\begin{center}
\includegraphics[width=75mm]{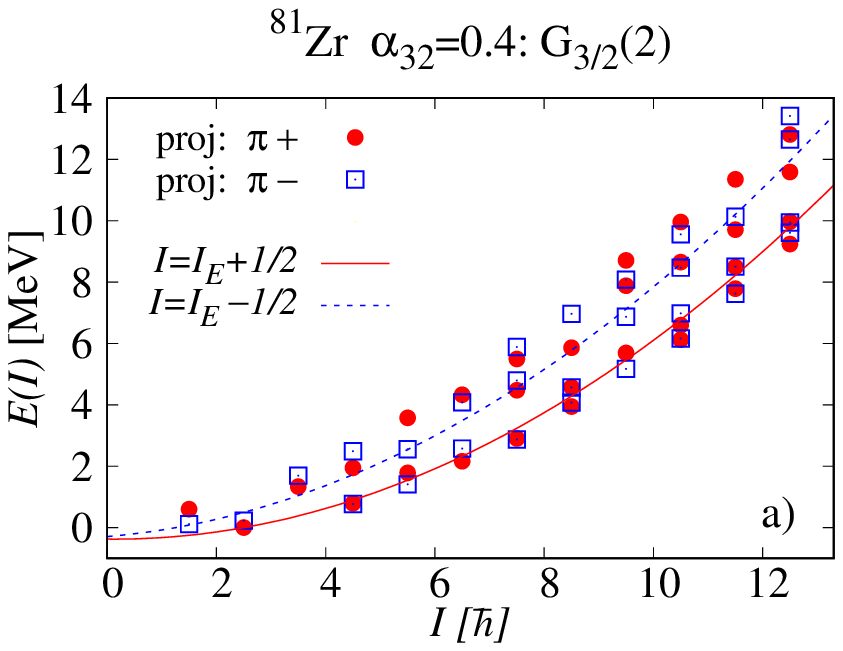}
\hspace*{-3mm}
\includegraphics[width=75mm]{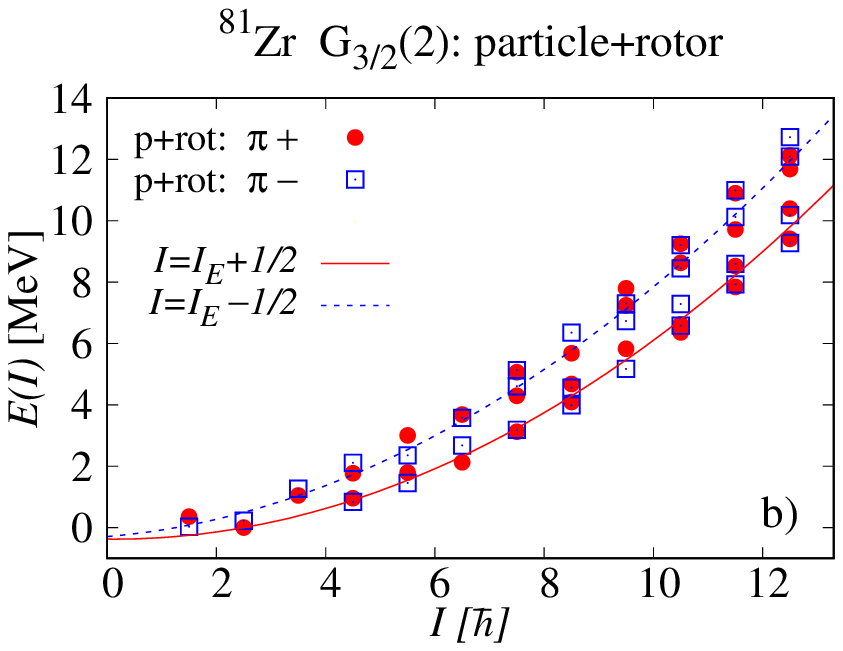}
\end{center}
\vspace*{-8mm}
\caption{(Color online)
Similar to Fig.~\ref{fig:G32a} but for
one neutron occupying a member of the second $G_{3/2}$ four-fold degenerate orbital
above the $N=40$ magic number.
}
\label{fig:G32b}
\end{figure}

\begin{figure}[!htb]
\begin{center}
\includegraphics[width=75mm]{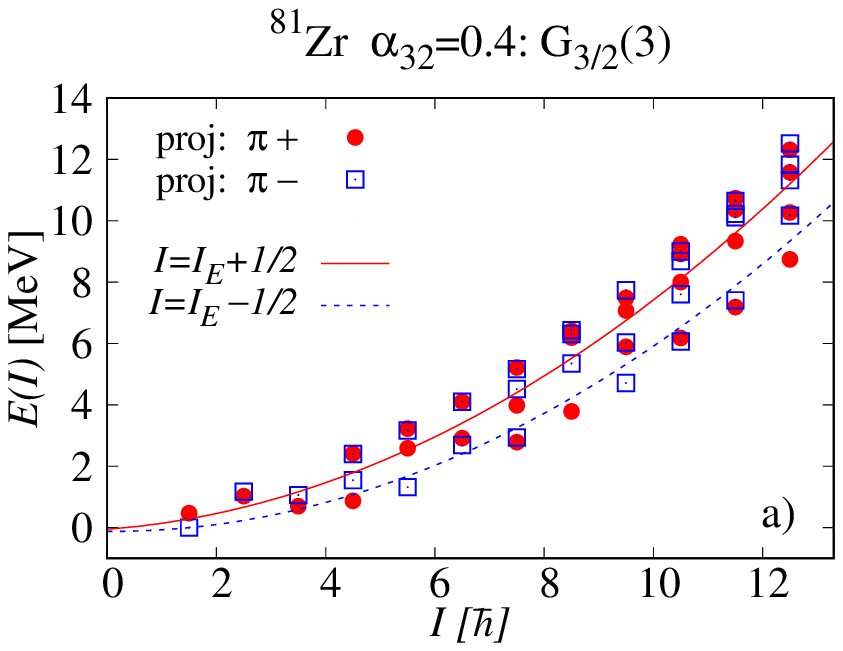}
\hspace*{-3mm}
\includegraphics[width=75mm]{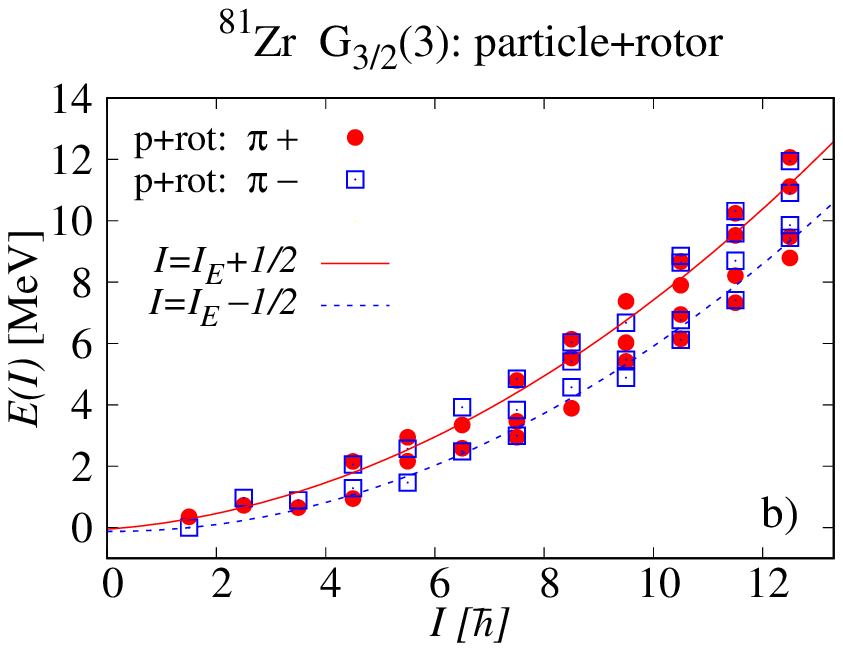}
\end{center}
\vspace*{-8mm}
\caption{(Color online)
Similar to Fig.~\ref{fig:G32a} but for
one neutron occupying a member of the third $G_{3/2}$ four-fold degenerate orbital
above the $N=40$ magic shell closure.
}
\label{fig:G32c}
\end{figure}

In the case of $G_{3/2}$ representation the numerical diagonalization of the Hamiltonian is
necessary for the exact solution taking into account the Coriolis coupling
as discussed in Sec.~\ref{sec:CoriG32},
although the Hamiltonian-matrix dimensions are small, $n_{I\pi}^{G_{3/2}} \le 4$
up to the highest spin $I=25/2$ shown in the present work.
In Figs.~\ref{fig:G32a}$-$\ref{fig:G32c},
the spectra of the microscopic projection calculation
and of the Coriolis-coupling model are presented
in the left and right panels, respectively, where one neutron occupies
the lowest three $G_{3/2}$ orbitals
(here the four degenerate orbitals are counted as one)
in the configurations illustrated in these figures.
The approximate energy expression in Eq.~(\ref{eq:EdecG32}),
where the second decoupling parameter is vanishing,
is also included as the solid and dotted lines in each figure.
The calculated values of the decoupling parameters and
of the moment of inertia are
\begin{align}
 &a^{G_{3/2}}_1= + 2.86,\ a^{G_{3/2}}_2=-0.47,
 \quad {\cal J}=7.48\ \mbox{[$\hbar^2$/MeV]},
\label{para:G32a} \\
 &a^{G_{3/2}}_1= + 1.87,\ a^{G_{3/2}}_2=-1.25,
 \quad {\cal J}=7.53\ \mbox{[$\hbar^2$/MeV]},
\label{para:G32b} \\
 &a^{G_{3/2}}_1=-1.75,\ a^{G_{3/2}}_2=-1.45,
 \quad {\cal J}=8.14\ \mbox{[$\hbar^2$/MeV]},
\label{para:G32c}
\end{align}
for the cases of the one-neutron occupying
the lowest, second and third $G_{3/2}$ orbital, respectively,
corresponding to Figs.~\ref{fig:G32a},~\ref{fig:G32b} and~\ref{fig:G32c}.

The energy spectra for the case of $G_{3/2}$ representation are much more complicated as compared to the two previously discussed cases and
the behavior of the splitting patterns is rather different from
those of $E_{1/2}$ and $E_{5/2}$ irreducible representations.
In particular, the spectrum is not necessarily composed of two sequences;
more sequences can be recognized in Figs.~\ref{fig:G32b} and~\ref{fig:G32c}.
Comparing the left and right panels
in Figs.~\ref{fig:G32a}$-$\ref{fig:G32c},
it can be seen that the agreement between the results of
the microscopic projection calculations and of the Coriolis-coupling model
is not so striking as in the cases of $E_{1/2}$ and $E_{5/2}$;
the splitting at each spin value is
slightly underestimated in the Coriolis-coupling model.
However, general patterns of the energy splitting seen in the microscopic
projection calculations are rather well-reproduced by the model.
For example, the energy ordering of the opposite parity states
at each spin value is reproduced correctly for most of the states.

An instructive example is provided by the result in Fig.~\ref{fig:G32a},
where the energy spectrum approximately splits into two sequences
similarly to the cases of $E_{1/2}$ and $E_{5/2}$.
This is because the second decoupling parameter $a^{G_{3/2}}_2$
is accidentally small as shown in Eq.~(\ref{para:G32a});
in such a case the simple analytic expression in Eq.~(\ref{eq:EdecG32})
is approximately valid and the spectrum follows
the expected pattern, although the correspondence is not perfect.
The results confirm that these two sequences are composed of
the spin-parity states of the $E$-representation shifted by spin $\pm 1/2$,
exactly as predicted by Eq.~(\ref{eq:EdecG32}).
In the other cases shown in Figs.~\ref{fig:G32b} and~\ref{fig:G32c},
where the second term of the Coriolis coupling in Eq.~(\ref{eq:G32CorCa})
is non-negligible, the resulting energy spectra are
more and more perturbed at increasing spins.
Although the approximate expression in Eq.~(\ref{eq:EdecG32}) gives a rough
estimate of the size of the splitting, the calculated energy splitting
is getting irregular yet centering around the lines given by Eq.~(\ref{eq:EdecG32}).
The differences in terms of energies between
the microscopic projection calculations and the Coriolis-splitting model
are larger.
This may indicate that the higher-order Coriolis-coupling
is more important in this case, or that the effects of the coupling
of an odd nucleon to other degrees of freedom appears to be more pronounced.
It is worth emphasizing that in the microscopic calculations,
there is no rotor-like contribution introduced when the residual interactions
between the constituent nucleons are diagonalized within
the angular-momentum and parity projection method.

%------------------------------------------------------------------------------
%------------------------------------------------------------------------------
\section{Summary and conclusions}
\label{sec:concl}

In the present work we have studied the effect of the tetrahedral symmetry
on deformed odd-mass nuclei employing the modeling in terms of the Coriolis coupling.
Limiting ourselves to the simplest example, we have restricted our considerations to
the tetrahedral doubly-magic core $Z=N=40$ plus one-particle systems.
For such quantum systems the eigenstates can be classified
by the irreducible representations of the point-group symmetry of the system.
For the tetrahedral double group $T_d^D$ there are three such representations,
$E_{1/2}$, $E_{5/2}$ and $G_{3/2}$, cf.~Ref.~\cite{Ham62}.
As it is well-known, for the axially-symmetric quadrupole-deformed nuclei
the Coriolis coupling makes a single $K=1/2$ rotational band
split into two sequences
for the spectra of the core plus one-particle systems.
We have calculated the matrix element of the Coriolis coupling
for the tetrahedrally-deformed case analytically.
It is found that the expression for the energy spectra contains
one parameter in the case of the $E_{1/2}$ and $E_{5/2}$ representations
and two parameters for the $G_{3/2}$ representation,
which are called generalized decoupling parameter(s)
and calculated uniquely by using the deformed intrinsic single-particle states.
The energy spectra of the $E_{1/2}$ and $E_{5/2}$ cases are shown to
split into two rotational bands like the case of the $K=1/2$ band
of the axially-symmetric nuclei.
The spectrum of the $G_{3/2}$ case is generally more complicated,
and it splits into two parabolic sequences
only when one of the two decoupling parameters vanishes.

In order to double-test the predicted properties of the rotational-energy spectra
for the tetrahedral-symmetric core plus one-particle systems, we have performed
the microscopic angular-momentum and parity projection calculations
for a prototype nucleus $^{81}$Zr.
Relatively large tetrahedral deformation was assumed to obtain
the well pronounced rotational bands.
The Woods-Saxon mean-field and the separable-type schematic interactions
have been employed following the approach of Refs.~\cite{TS12,TSD13}.
By occupying the proper single-particle state above the $N=Z=40$
tetrahedral shell-closure, the resulting energy spectra
corresponding to the $E_{1/2}$, $E_{5/2}$ and $G_{3/2}$ representations
are obtained. We found that these spectra can be well
reproduced by the energy expressions resulting from the Coriolis coupling
for the $E_{1/2}$ and $E_{5/2}$ representations.
While the level-to-level correspondence is not so direct as the above two cases,
the spectra for the $G_{3/2}$ representation can be approximately represented using the generalized decoupling parameter concepts.
It should be emphasized that no rotor-type structure is assumed
in the microscopic projection calculations.
The illustrated correspondence between the results of the tetrahedral Coriolis-coupling model and microscopic spin and parity projected calculations
suggests that the picture of tetrahedral nuclear rotor
can be well justified at least for the large deformation.

It is worthwhile noticing that the tetrahedral equilibrium deformations
predicted by microscopic calculations are always lower than the 0.4 value
taken here for an illustration of the asymptotic regime.
For example, realistic potential energy calculations
give as minimum deformations $\alpha_{32}\approx 0.2$ for $^{80}$Zr,
see e.g., Fig.~1 of Ref.~\cite{TSF14}; the values predicted for
the tetrahedral equilibria in other nuclei are similar or smaller.
For the tetrahedral deformations in the vicinity of  $\alpha_{32}\approx 0.2$
the spectrum has neither clearly parabolic nor clearly linear spin dependence
for the $A_1$ representation of the core nucleus~\cite{TSD13},
and the eigenstates do not follow any single (approximately)
parabolic sequence but rather scatter around it. In such a case,
the core plus one-particle system shows more complicated spectrum,
because additional contributions caused by the Coriolis-coupling
are generally non-negligible.
However, the importance of the present approach lies in providing
a relatively simple asymptotic-limit description of the impact of
an odd nucleon on the core nucleus that shows ideal rotational spectrum.
At the same time it illustrates the practical applications
of the group theoretical considerations as a very powerful tool
when studying the nuclear point-group symmetries.

In the present work, we have studied the rotational spectrum
of relatively simple cases of the core plus one-particle systems
with large tetrahedral deformation, from the point of view of the structure of the Coriolis coupling. We expect that the present study will contribute to
investigation and deeper understanding of the general case of
nuclear structure under the tetrahedral symmetry, in particular via establishing the asymptotic properties of the Coriolis-coupling term at the strong-coupling limit.

%------------------------------------------------------------------------------
%------------------------------------------------------------------------------
\appendix

%------------------------------------------------------------------------------
\section{Doublex eigenvalues for various irreducible representations and
simplex symmetry}
\label{sec:DblxSy}

In this Appendix, we briefly comment on
how the results in Tables~\ref{tab:dblx} and~\ref{tab:Sysym} are obtained.
Firstly, the values of the $z$-doublex for each irreducible representation
are found using Eq.~(\ref{eq:jmdblx}) for the states
with the lowest angular-momentum and parity allowed for it.
For the $A_1$ representation, the lowest state has $I^\pi=0^+$ and
then trivially $\mu=0$. Since $A_2$ is parity-conjugate to $A_1$ it follows that
$\mu(A_2)=\mu(A_1)+2 \equiv 2$~(mod~4).
For $F_1$ the lowest state has $I^\pi=1^+$, i.e. the possible $K$-values are $K=0,\pm 1$,
therefore, $\mu=0,\pm 1$ in this case. Similarly, $F_2$ is parity-conjugate to $F_1$ and consequently
$\mu(F_2)=\mu(F_1)+2 \equiv 2,\mp 1$~(mod~4).
In the case of $E$-representation, the lowest state has $I^\pi=2^+$, i.e. $K=0,\pm 1,\pm 2$,
but $I^\pi=2^+$ also appears for $F_2$, which has $\mu=\pm 1,2$,
so $E$ should have the remaining $\mu=0,2$ (note that $-2 \equiv 2$~(mod~4)).
For $E_{1/2}$ the lowest state has $I^\pi=1/2^+$, what implies $\mu=\pm 1/2$.
$E_{5/2}$ is parity-conjugate to $E_{1/2}$, and thus
$\mu(E_{1/2})=\mu(E_{5/2})+2 \equiv \mp 3/2$~(mod~4).
For $G_{3/2}$ the lowest state has $I^\pi=3/2^+$,
and then $\mu=\pm 1/2,\pm 3/2$.
This completes the discussion of the content of Table~\ref{tab:dblx}.

The simplex symmetry (related to the operator $\hat S_y$ in the text)
can similarly be found by considering
the lowest possible angular-momentum and parity states with $K=0$ because
of the following relation:
\begin{align}
 \hat S_y |I^\pi K \rangle=\pi\,(-1)^{I+K}|I^\pi -K \rangle.
\label{eq:IKSy}
\end{align}
Thus, for $\lambda\mu=A_1 0$, which has $I^\pi=0^+$,
eigenvalue $s$ of $\hat S_y$ is $s=+1$.
For $A_2 2$, which has $I^\pi=0^-$, we obtain $s=-1$.
For $F_1 0$, which has $I^\pi=1^+$, the eigenvalue of interest is $s=-1$,
while for $F_2 2$, which has $I^\pi=1^-$, we find $s=+1$.
As for $E 0$, $I^\pi=2^+$, and thus $s=+1$,
while for $E 2$, $I^\pi=2^-$, and consequently  $s=-1$.
These results are summarized in Table~\ref{tab:Sysym}.

%------------------------------------------------------------------------------
\section{Auxiliary identities for
expansion coefficients within  $\mbold{A_1}$, $\mbold{A_2}$ and $\mbold{E}$ representations}
\label{sec:ApPropwf}

In order to calculate the Coriolis coupling
for the core plus one-particle systems within the irreducible-representation bases,
we need some auxiliary identities for the expansion coefficients in Eq.~(\ref{eq:Ccoef})
of the $A_1$, $A_2$ and $E$ representations.
They are derived from the following expression valid for
the basis states $|I^\pi\lambda\mu\beta\rangle$
of an arbitrary representation $\lambda$,
\begin{align}
 &\sum_{\mu}\langle I^\pi\lambda\mu\beta|\hat D^\dagger(g) \hat O \hat D(g)
  |I^\pi\lambda\mu\beta\rangle
                                                      \nonumber \\
 &\quad=\sum_{\mu\mu'\mu''}
  D_{\mu''\mu}^{[\lambda] *}(g)\langle I^\pi\lambda\mu''\beta| \hat O
  |I^\pi\lambda\mu'\beta\rangle D_{\mu'\mu}^{[\lambda]}(g)
 =\sum_{\mu}\langle I^\pi\lambda\mu\beta|\hat O |I^\pi\lambda\mu\beta\rangle \,,
                                                      \label{eq:exOtr}
\end{align}
$\forall\; g \in G$. Above $\hat O$ denotes an arbitrary operator. In obtaining this result the unitarity of the representation matrix
$D_{\mu\mu'}^{[\lambda]}(g)$ has been used.
Applying Eq.~(\ref{eq:exOtr})
with $\hat{D}(g)=\hat S_y$ and $\hat O=\hat J_z$
and using the fact that $\hat S_y^\dagger \hat J_z \hat S_y=-\hat J_z$,
one can demonstrate that
\begin{align}
       \sum_{\mu}\langle I^\pi\lambda\mu\beta|\hat J_z | I^\pi\lambda\mu\beta\rangle=0.
                                                    \label{eq:form1}
\end{align}
Next, we note that the group element $\hat S_4$ in the class $S_4$,
\begin{align}
 \hat S_4 \equiv \hat \Pi\, e^{i\frac{\pi}{4}\hat J_z}
 e^{i\frac{\pi}{2}\hat J_y} e^{-i\frac{\pi}{4}\hat J_z},
\label{eq:S4xp}
\end{align}
transforms the operator $\hat J_z$ as
\begin{align}
 \hat S_4^\dagger \hat J_z \hat S_4
 =\frac{1}{\sqrt{2}}(\hat J_x - \hat J_y)\equiv \hat J_{x'} \,.
\label{eq:S4xpJz}
\end{align}
In full analogy, the group element $\hat S'_4$,
\begin{align}
 \hat S'_4 \equiv \hat \Pi\, e^{-i\frac{\pi}{4}\hat J_z}
 e^{-i\frac{\pi}{2}\hat J_y} e^{i\frac{\pi}{4}\hat J_z},
\label{eq:S4yp}
\end{align}
transforms the operator $\hat J_z$ as follows
\begin{align}
 \hat {S'_4}^{\dagger} \hat J_z \hat S'_4
 =-\frac{1}{\sqrt{2}}(\hat J_x + \hat J_y)\equiv \hat J_{y'} \,.
\label{eq:S4ypJz}
\end{align}
Applying Eq.~(\ref{eq:exOtr})
with $\hat{D}(g)=\hat S_4$ and $\hat O=\hat J_z^2$
and with $\hat{D}(g)=\hat S'_4$ and $\hat O=\hat J_z^2$, one obtains
\begin{align}
  &\sum_\mu \langle I^\pi\lambda\mu\beta|\hat J_z^2 | I^\pi\lambda\mu\beta\rangle
 =\sum_\mu \langle I^\pi\lambda\mu\beta|\hat J_{x'}^2 | I^\pi\lambda\mu\beta\rangle
 =\sum_\mu \langle I^\pi\lambda\mu\beta|\hat J_{y'}^2 | I^\pi\lambda\mu\beta\rangle,
\label{eq:Oj2exp}
\end{align}
and consequently,
\begin{align}
 \sum_{\mu} \langle I^\pi\lambda\mu\beta|\hat J_z^2 | I^\pi\lambda\mu\beta\rangle
 =\frac{1}{3}\sum_\mu \langle I^\pi\lambda\mu\beta|\mbold{\hat J}^2 | I^\pi\lambda\mu\beta\rangle
= \frac{f_{\lambda}}{3}\,I(I+1).
\label{eq:form2}
\end{align}

Now, let us consider the $A_1$ expansion coefficients.
We use the symbols $(k,m)$ instead of $(I,K)$ for integer angular-momentum;
$(I,K)$ is kept for the half-odd integer angular-momentum in odd nuclei
considered in the next Appendix.
From the $\hat S_y$ symmetry relations in Table~\ref{tab:Sysym}, we deduce
\begin{align}
       C^{\pi}_{km,A_10\beta}=\pi (-1)^{k+m}C^{\pi}_{k-m,A_10\beta} \,.
\label{eq:SyA1}
\end{align}
The following identities can be derived by using the normalization condition
and Eqs.~(\ref{eq:form1}) and~(\ref{eq:form2}),
\begin{align}
 &\sum_m |C^{\pi}_{km,A_10\beta}|^2 = 1,
\label{eq:CA1f0} \\
 &\sum_m |C^{\pi}_{km,A_10\beta}|^2 m = 0,
\label{eq:CA1f1} \\
 &\sum_m |C^{\pi}_{km,A_10\beta}|^2 m^2 = \frac{1}{3}k(k+1).
\label{eq:CA1f2}
\end{align}
The $A_2$ representation is parity-conjugate of the $A_1$ representation,
and therefore $C^{\pi}_{km,A_22\beta}= C^{(-\pi)}_{km,A_10\beta}$,
from which the following identities can be derived:
\begin{align}
 C^{\pi}_{km,A_22\beta}=-\pi (-1)^{k+m}C^{\pi}_{k-m,A_22\beta} \,,
\label{eq:SyA2}
\end{align}
together with
\begin{align}
 &\sum_m |C^{\pi}_{km,A_22\beta}|^2 = 1,
\label{eq:CA2f0} \\
 &\sum_m |C^{\pi}_{km,A_22\beta}|^2 m = 0,
\label{eq:CA2f1} \\
 &\sum_m |C^{\pi}_{km,A_22\beta}|^2 m^2 = \frac{1}{3}k(k+1).
\label{eq:CA2f2}
\end{align}
For the $E$ expansion coefficients ($f_{E}=2$),
the $\hat S_y$ symmetry relations in Table~\ref{tab:Sysym} give
\begin{align}
 C^{\pi}_{km,E\mu\beta}=(-1)^{\mu/2}\,\pi (-1)^{k+m}C^{\pi}_{k-m,E\mu\beta}\,,
 \quad (\mu=0,2),
\label{eq:SyE}
\end{align}
and the normalization conditions and
Eqs.~(\ref{eq:form1}) and~(\ref{eq:form2}) lead to
\begin{align}
 &\sum_{\mu=0,2}\sum_m |C^{\pi}_{km,E\mu\beta}|^2 = 2,
\label{eq:CEf0} \\
 &\sum_{\mu=0,2}\sum_m |C^{\pi}_{km,E\mu\beta}|^2 m = 0,
\label{eq:CEf1} \\
 &\sum_{\mu=0,2}\sum_m |C^{\pi}_{km,E\mu\beta}|^2 m^2 = \frac{2}{3}k(k+1).
\label{eq:CEf2}
\end{align}

%------------------------------------------------------------------------------
\section{Detailed evaluation of the Coriolis-coupling matrix element}
\label{sec:ApEvC}

In this Appendix we calculate the diagonal matrix elements
in Eq.~(\ref{eq:CorIE12r}) for $E_{1/2}$,
in Eq.~(\ref{eq:CorIE52r}) for $E_{5/2}$,
and in Eqs.~(\ref{eq:G32relME1}) and~(\ref{eq:CorIG32r}) for $G_{3/2}$,
by using the properties of the expansion coefficients discussed
in Appendix~\ref{sec:ApPropwf}.

First consider the case of $E_{1/2}$.
As it is discussed in Sec.~\ref{sec:CoriE12},
the basis state for $(\lambda\mu)\equiv(E_{1/2}1/2)$ is constructed by
\begin{align}
 {\cal N}^{I\pi}_{\lambda \mu \alpha}| I^\pi \lambda \mu \alpha \rangle
 & \equiv \Bigl[ | k^{\pi} A_1 0 \gamma \rangle
 \otimes | \mhalf^+ \mhalf \rangle \Bigr]_I
 \nonumber \\
 &= \sum_{m}
 | I^\pi K \rangle  C^{\pi}_{k m,A_10\gamma}
 \langle k m \mhalf\mhalf | I K \rangle ,
\end{align}
where $\alpha=(k\gamma)$ with $k=I\pm \mhalf$ and $K=m+\mhalf$,
and ${\cal N}^{I\pi}_{\lambda \mu \alpha}$ is normalization constant of $| I^\pi \lambda \mu \alpha \rangle$.

For the case of $I=k+\mhalf$,
by inserting the expression of the Clebsch-Gordan coefficient,
the result can be written explicitly,
\begin{align}
  {\cal N}^{I\pi}_{\lambda \mu \alpha} | I^\pi \lambda \mu \alpha \rangle
 &= \sum_m  (-1)^{k-\frac{1}{2}+I}
 | I^\pi K \rangle  C^{\pi}_{k m,A_10\gamma}
   \sqrt{\mathstrut \frac{k+m+1}{2k+1}},
\end{align}
and then the normalization constant can be calculated by
using the identities in Eqs.~(\ref{eq:CA1f0}) and~(\ref{eq:CA1f1}),
\begin{align}
 &|{\cal N}^{I\pi}_{\lambda \mu \alpha}|^2
 = \sum_m  |C^{\pi}_{k m,A_10\gamma}|^2
  \frac{k+m+1}{2k+1} = \frac{k+1}{2k+1},
 \nonumber \\
 & \Rightarrow\quad {\cal N}^{I\pi}_{\lambda \mu \alpha}
  \equiv  (-1)^{k-\frac{1}{2}+I} \sqrt{\mathstrut \frac{k+1}{2k+1}}.
\label{eq:Normex}
\end{align}
Thus,
\begin{align}
 | I^\pi \lambda \mu  \alpha \rangle
 &= \sum_m | I^\pi K \rangle  C^{\pi}_{k m,A_10\gamma}
  \sqrt{\mathstrut \frac{k + m+1}{k+1}}
  = \sum_m | I^\pi K \rangle  C^{\pi}_{k m,A_10\gamma}
  \sqrt{\mathstrut \frac{I+K}{k+1}}\,.
\label{eq:E12Ikpwf}
\end{align}
With this wave function, its simplex conjugate state can be obtained as follows
\begin{align}
 | I^\pi \lambda-\mu \alpha \rangle
 &\equiv  \hat S_y | I^\pi \lambda \mu \alpha \rangle \nonumber \\
 &= \sum_m \pi (-1)^{I+K}
   | I^\pi -K \rangle  C^{\pi}_{k m,A_10\gamma}
   \sqrt{\mathstrut \frac{I+K}{k+1}} \nonumber \\
 &= \sum_m \pi (-1)^{k-m+1}
   | I^\pi K-1 \rangle  C^{\pi}_{k -m,A_10\gamma}
   \sqrt{\mathstrut \frac{I-K+1}{k+1}} \nonumber\\
 &= - \sum_m
   | I^\pi K-1 \rangle  C^{\pi}_{k m,A_10\gamma}
   \sqrt{\mathstrut \frac{I-K+1}{k+1}},
\end{align}
where the property of the $A_1$ coefficient in Eq.~(\ref{eq:SyA1}) is used.
Then, the collective diagonal matrix-elements displayed below can be calculated as
\begin{align}
 \langle I^\pi \lambda \mu \alpha | \hat I_3 | I^\pi \lambda \mu \alpha\rangle
 &= \sum_m | C^{\pi}_{k m,A_10\gamma}|^2
 \frac{k +m+1}{k+1}\left(m +\frac{1}{2}\right)
  = \frac{1}{3}k+\frac{1}{2}
  =\frac{1}{3}(I+1),
\end{align}
and
\begin{align}
 \langle I^\pi \lambda \mu \alpha | \hat I_- \hat S_y
  | I^\pi \lambda\mu \alpha \rangle
 & =  \sum_K \langle I^\pi \lambda +\mu \alpha |I K \rangle
 \langle IK | \hat I_- | I^\pi \lambda -\mu \alpha \rangle \nonumber \\
 &=  \sum_K \langle I^\pi \lambda +\mu \alpha |I K \rangle
  \sqrt{\mathstrut (I+K)(I-K+1)} \langle IK-1| I^\pi \lambda -\mu \alpha \rangle
 \nonumber\\
 &= - \sum_m |  C^{\pi}_{k m,A_10\gamma}|^2 \sqrt{\mathstrut \frac{I+K}{k+1}}
  \sqrt{\mathstrut (I+K)(I-K+1)} \sqrt{\mathstrut \frac{I-K+1}{k+1}}
 \nonumber \\
 &= - \sum_m | C^{\pi}_{k m,A_10\gamma}|^2 \frac{(k+m+1)(I-K+1)}{k+1}
 \nonumber \\
 &= - \bigl((I+1) -\langle I^\pi \lambda \mu \alpha | \hat I_3
 | I^\pi \lambda \mu \alpha \rangle \bigr)
 =- 2 \langle I^\pi \lambda \mu \alpha | \hat I_3
 | I^\pi \lambda \mu \alpha \rangle,
\end{align}
where the identities in Eqs.~(\ref{eq:CA1f0})$-$(\ref{eq:CA1f2}) have been used.

For the case of $I=k-\mhalf$,
\begin{align}
 {\cal N}^{I\pi}_{\lambda \mu \alpha} | I^\pi \lambda \mu \alpha \rangle
 &= (-1)^{k-\frac{1}{2}+I} \sum_m
 | I^\pi K \rangle  C^{\pi}_{k m,A_10\gamma}
   \sqrt{\mathstrut \frac{k-m}{2k+1}},
\end{align}
and the normalization constant can be calculated similarly as before as,
\begin{align}
 &|{\cal N}^{I\pi}_{\lambda \mu \alpha}|^2 = \sum_m  |C^{\pi}_{k m,A_10\gamma}|^2
   \frac{k-m}{2k+1}=\frac{k}{2k+1},
 \nonumber \\
 & \Rightarrow\quad {\cal N}^{I\pi}_{\lambda \mu \alpha}
   \equiv (-1)^{k-\frac{1}{2}+I}
  \sqrt{\mathstrut \frac{k}{2k+1}},
\end{align}
and it follows that
\begin{align}
 | I^\pi \lambda \mu \alpha \rangle
 &= \sum_m | I^\pi K \rangle  C^{\pi}_{k m,A_10\gamma}
   \sqrt{\mathstrut \frac{k - m}{k}}
  = \sum_m | I^\pi K \rangle  C^{\pi}_{k m,A_10\gamma}
   \sqrt{\mathstrut \frac{I-K+1}{k}}\,.
\end{align}
Its simplex conjugate state can be written as
\begin{align}
 | I^\pi \lambda-\mu \alpha \rangle
 &\equiv \hat S_y | I^\pi \lambda \mu \alpha \rangle \nonumber \\
 &= \sum_m \pi (-1)^{I+K}
   | I^\pi -K \rangle  C^{\pi}_{k m,A_10\gamma}
   \sqrt{\mathstrut \frac{I-K+1}{k}} \nonumber \\
 &= \sum_m \pi (-1)^{k-m}
   | I^\pi K-1 \rangle  C^{\pi}_{k -m,A_10\gamma}
   \sqrt{\mathstrut \frac{I+K}{k}} \nonumber \\
 &= \sum_m \
   | I^\pi K-1 \rangle  C^{\pi}_{k m,A_10\gamma}
   \sqrt{\mathstrut \frac{I+K}{k}} \,,
\end{align}
and the corresponding diagonal matrix elements are
\begin{align}
 \langle I^\pi \lambda \mu \alpha | \hat I_3 | I^\pi \lambda \mu \alpha\rangle
 &= \sum_m | C^{\pi \cdot \pi_0}_{k m,A_10\gamma}|^2
 \frac{k -m}{k} \left(m +\frac{1}{2}\right)
  = -\frac{1}{3}(k+1)+\frac{1}{2}
  =-\frac{1}{3}I,
\end{align}
and
\begin{align}
 \langle I^\pi \lambda \mu \alpha | \hat I_-\hat S_y
 | I^\pi \lambda \mu \alpha \rangle
  & =  \sum_K \langle I^\pi \lambda +\mu \alpha |I K \rangle \langle IK |
 \hat I_- | I^\pi \lambda -\mu \alpha \rangle \nonumber \\
  &=  \sum_K \langle I^\pi \lambda +\mu \alpha |I K \rangle
  \sqrt{\mathstrut (I+K)(I-K+1)} \langle IK-1| I^\pi \lambda -\mu \alpha \rangle
    \nonumber \\
  &= \sum_m |  C^{\pi}_{k m,A_10\gamma}|^2
  \sqrt{\mathstrut \frac{I-K+1}{k}}
  \sqrt{\mathstrut (I+K)(I-K+1)} \sqrt{\mathstrut \frac{I+K}{k}} \nonumber \\
  &= \sum_m | C^{\pi}_{k m,A_10\gamma}|^2 \frac{(k-m)(I+K)}{k}
    \nonumber \\
  &= I+\langle I^\pi \lambda \mu \alpha | \hat I_3
  | I^\pi \lambda \mu \alpha \rangle
  =- 2 \langle I^\pi \lambda \mu \alpha | \hat I_3
  | I^\pi \lambda \mu \alpha \rangle.
\end{align}
In this way the validity of Eq.~(\ref{eq:CorIE12r}) for $E_{1/2}$ is demonstrated.

As for Eq.~(\ref{eq:CorIE52r}) for $E_{5/2}$, the same calculation
can be repeated with replacing the $A_1$ expansion coefficient
$C^{\pi}_{k m,A_10\gamma}$ by the $A_2$ coefficient
$C^{\pi}_{k m,A_22\gamma}$ and
employing Eqs.~(\ref{eq:SyA2})$-$(\ref{eq:CA2f2}),
the only difference being the sign of the matrix element
of $\hat I_- \hat S_y$.

For the case of $G_{3/2}$, for example with $I=k+\mhalf$,
employing the wave functions in Eq.~(\ref{eq:G32mbasis}),
one can derive
\begin{align}
 &\langle I G_{3/2} 1/2 \alpha|\hat I_3 | I G_{3/2} 1/2 \alpha \rangle
 +\langle I G_{3/2} -3/2 \alpha|\hat I_3 | I G_{3/2} -3/2 \alpha \rangle
 \nonumber \\
 &\quad = \sum_{\mu=0,2}\sum_m | C^{\pi}_{k m,E\mu\gamma}|^2
 \frac{k +m+1}{k+1}\left(m +\frac{1}{2}\right)
  =\frac{2}{3}(I+1),
\end{align}
and
\begin{align}
 &-\langle I G_{3/2} 1/2 \alpha|\hat I_- \hat S_y |I G_{3/2} 1/2 \alpha \rangle
 +\langle I G_{3/2} -3/2 \alpha|\hat I_- \hat S_y |I G_{3/2} -3/2 \alpha \rangle
 \nonumber \\
 &= \sum_{\mu=0,2}\sum_m
 | C^{\pi}_{k m,E\mu\gamma}|^2 \frac{(k+m+1)(I-K+1)}{k+1}
 =\frac{4}{3}(I+1)
 \nonumber \\
 &=2\bigl(
  \langle I G_{3/2} 1/2 \alpha|\hat I_3 | I G_{3/2} 1/2 \alpha \rangle
 +\langle I G_{3/2} -3/2 \alpha|\hat I_3 | I G_{3/2} -3/2 \alpha \rangle
 \bigr),
\end{align}
where Eqs.~(\ref{eq:SyE})$-$(\ref{eq:CEf2}) are used:
The calculation is similar with $I=k-\mhalf$.
In this way, the diagonal matrix elements of the Coriolis coupling
can be evaluated analytically for all the three representations.

As stated in the text (Sec.~\ref{sec:CoriE12}),
the fact that the non-diagonal matrix elements of
$\hat I_3$ and $\hat I_- \hat S_y$ vanish
in these basis states is confirmed by numerical calculations.
It is also confirmed that the eigenvalues of
$\hat I_3$ and $\hat I_- \hat S_y$ coincide with
the calculated diagonal matrix elements above,
which suggests that these specific basis states indeed diagonalize
$\hat I_3$ and $\hat I_- \hat S_y$ operators.

%------------------------------------------------------------------------------
\section{Relation between $\mbold{\mu=1/2}$ and $\mbold{\mu=-3/2}$
         basis-states for the $\mbold{G_{3/2}}$ representation}
\label{sec:ApBsG32}

Operations of the group elements transform the basis states
within each irreducible representation.
It follows that basis state $\mu=-3/2$ of $G_{3/2}$ can be obtained
from the $\mu=1/2$ state by
\begin{align}
 | G_{3/2} -3/2\rangle = \hat X_+| G_{3/2} 1/2\rangle ,
\label{eq:G32X1}
\end{align}
with the suitably chosen operator $\hat X_+$,
which is a linear combination of the group elements of $T_d^D$.
Operator $\hat S_4$ defined in Eq.~(\ref{eq:S4xp}) can be used
because $(\hat S^{}_4+\hat S^\dagger_4)$ conserves the $z$-signature,
\begin{align}
 \hat R^{\dagger}_z(\hat S^{}_4+\hat S^\dagger_4)\hat R_z
 =(\hat S^{\dagger}_4+\hat S_4),
 \quad
 \hat R_z\equiv e^{i\pi \hat J_z},
\label{eq:S4Sz}
\end{align}
and therefore it transforms the $\mu=1/2$ state
within the two-dimensional subspace spanned by $\mu=1/2$ and $\mu=-3/2$ states
(note that the $z$-doublex $\mu=1/2,-3/2$ states belong to
the $z$-signature $r=+1/2$ states,
and $\mu=-1/2,\,3/2$ states to $r=-1/2$ states).
Then, $\hat X_+ \propto (\hat S^{}_4+\hat S^\dagger_4+\xi)$
satisfies Eq.~(\ref{eq:G32X1}), where the constant $\xi$ is determined by
the condition $\langle G_{3/2} 1/2 |\hat X_+| G_{3/2} 1/2\rangle =0$
for the state with the lowest allowed spin $I=\m3half$
in the $G_{3/2}$ representation; one finds
$\xi=-2\langle \m3half\mhalf|e^{i\frac{\pi}{2}\hat J_y}| \m3half\mhalf\rangle
=1/\sqrt{2}$.
Taking into account of the normalization condition
$\langle G_{3/2} 1/2 |\hat X_+^{\dagger}\hat X_+| G_{3/2} 1/2\rangle =1$,
we find
\begin{align}
 \hat X_+\equiv i\sqrt{\frac{2}{3}}\,\Bigl(\hat S^{}_4+ \hat S^{\dagger}_4
 +\frac{1}{\sqrt{2}}\Bigr),
\label{eq:G32Xop1}
\end{align}
where
$\langle G_{3/2} 1/2 |(\hat S_4)^2| G_{3/2} 1/2\rangle
=\langle G_{3/2} 1/2 |(\hat S_4^{\dagger})^2| G_{3/2} 1/2\rangle=0$ is used.
It may be worthwhile noticing that the inverse relation
to Eq.~(\ref{eq:G32X1}) implies
\begin{align}
 | G_{3/2} 1/2\rangle = \hat X^{\dagger}_+| G_{3/2} -3/2\rangle .
\label{eq:G32X2}
\end{align}
In the same way, the $\mu=2$ basis state of the $E$ representation
can be obtained from the $\mu=0$ state,
\begin{align}
 | E2 \rangle = \hat X^E_+| E0\rangle ,
\label{eq:EXE}
\end{align}
with
\begin{align}
 \hat X^E_+\equiv i\sqrt{\frac{1}{3}}\,
 \Bigl(\hat S^{}_4+ \hat S^{\dagger}_4 +1 \Bigr),
\label{eq:EXEop}
\end{align}
which can be used for constructing the basis states of the $E$ representation.

%-------------------------------------------------------------------------------
%-------------------------------------------------------------------------------

\begin{acknowledgments}
One of the authors (J.D.) wishes to acknowledge a partial support from the Polish National Science Centre under Contract No.~2016/21/B/ST2/01227.
\end{acknowledgments}

\vspace*{10mm}

%-------------------------------------------------------------------------------

\end{document}